\documentclass[10pt,letterpaper]{article}
\usepackage[top=1in,footskip=0.75in,marginparwidth=2in]{geometry}

\usepackage[utf8]{inputenc}

\usepackage{cite}

\usepackage{nameref,hyperref}

\usepackage[right]{lineno}

\usepackage{microtype}
\DisableLigatures[f]{encoding = *, family = * }

\raggedright
\usepackage{changepage}

\usepackage[aboveskip=1pt,labelfont=bf,labelsep=period,singlelinecheck=off]{caption}

\makeatletter
\renewcommand{\@biblabel}[1]{\quad#1.}
\makeatother

\usepackage{lastpage,fancyhdr,graphicx}
\usepackage{epstopdf}
\fancyheadoffset[L]{0.25in}
\fancyfootoffset[L]{0.25in}
\fancyheadoffset[R]{0.25in}
\fancyfootoffset[R]{0.25in}

\usepackage{color}

\definecolor{Gray}{gray}{.25}

\usepackage{graphicx}

\usepackage{hyperref}
\usepackage{verbatim}
\usepackage[ruled,vlined,linesnumbered]{algorithm2e}
\usepackage{booktabs} 
\usepackage{caption}
\usepackage{threeparttable}
\usepackage{longtable} 
\usepackage{pdflscape}
\usepackage{rotating}
\usepackage{adjustbox}

\newcommand{\orsVarName}[1]{{\fontfamily{pcr}\selectfont #1}}
\newcommand{\sectionref}[1]{\S\ref{#1}}
\newcommand{\algref}[1]{Alg. \ref{#1}}
\newcommand{\figref}[1]{Fig. \ref{#1}}
\newcommand{\tabref}[1]{Table \ref{#1}}
\newcommand{\appref}[1]{Appendix \ref{#1}}

\begin{document}
\vspace*{0.35in}

\begin{flushleft}
{\Large
\textbf\newline{Imputing Missing Values in the Occupational Requirements Survey}
}
\newline
\\
Terry Leitch\textsuperscript{1,2},
Debjani Saha\textsuperscript{2}*
\\
\bigskip
\bf{1} Carey Business School - Johns Hopkins University
\\
\bf{2} ruxton.ai
\\
\bigskip
* debjani@ruxton.ai

\end{flushleft}


\section{Abstract}

The U.S. Bureau of Labor Statistics allows public access to much of the data acquired through its Occupational Requirements Survey (ORS). This data can be used to draw inferences about the requirements of various jobs and job classes within the United States workforce. However, the dataset contains a multitude of missing observations and estimates, which somewhat limits its utility. Here, we propose a method by which to impute these missing values that leverages many of the inherent features present in the survey data, such as known population limit and correlations between occupations and tasks. An iterative regression fit, implemented with a recent version of XGBoost and executed across a set of simulated values drawn from the distribution described by the known values and their standard deviations reported in the survey, is the approach used to arrive at a distribution of predicted values for each missing estimate. This allows us to calculate a mean prediction and bound said estimate with a 95\% confidence interval. 
We discuss the use of our method and how the resulting imputations can be utilized to inform and pursue future areas of study stemming from the data collected in the ORS. \footnote{A software package implementing our method and further downstream analyses is available on Github at \url{https://github.com/saharaja/imputeORS}.}
Finally, we conclude with an outline of WIGEM, a generalized version of our weighted, iterative imputation algorithm that could be applied to other contexts.

\section{Introduction} \label{sec:intro}
\subsection{Occupational Requirements Survey} \label{subsec:ors}

The Bureau of Labor Statistics (BLS) collects numerous data to capture important information regarding the workforce in the United States. One of the tools they use to do this is the Occupational Requirements Survey (ORS) \cite{ORSoverview}, a comprehensive survey that gathers information regarding the various demands and conditions (requirements) of different occupations within the U.S. economy. These requirements fall into one of four categories: physical demands; environmental conditions; education, training, and experience; and cognitive and mental requirements.
The initial data that were analyzed were publicly available datasets known as ``Wave 1'' and ``Wave 2.'' 
The survey is accomplished in repeated waves across several years; the data analyzed from the public set constituted Wave 1, year 2 and Wave 2, year 1. In this paper, we focus on the Wave 1 data.

\subsection{Structure of ORS Data} \label{subsec:struct}

The data used in this analysis are publicly available from the BLS  ORS, which is administered by the BLS to gather information regarding job-related demands and conditions across different occupations and labor categories. Importantly, this survey captures the distribution for the \emph{minimum} job requirements of various occupations, rather than the \emph{actual qualifications} of individuals working in these occupations (i.e., the survey does not capture information of under- and/or over-qualified employees). Additionally, this dataset covers only civilian occupations.

Numerous types of data are captured in the ORS so we focused on the information we identified as relevant to our goal of imputing missing estimates for population distributions of occupational requirements. For any given observation, the relevant fields utilized from the ORS were as follows:

\begin{enumerate}
	\setlength\itemsep{0em}
	\item \orsVarName{\textbf{upper\_soc\_code}} - Numeric field \\Classifies occupation using the Standard Occupational Classification (SOC) system \cite{SOCguide}
	\item \orsVarName{\textbf{occupation\_text}} - Character field \\Descriptive text of occupation
	\item \orsVarName{\textbf{data\_element\_text}} - Character field \\Descriptive text of job requirement/demand 
	\item \orsVarName{\textbf{data\_type\_text}} - Character field \\Descriptive text of levels associated with any given requirement/demand
	\item \orsVarName{\textbf{estimate\_type\_text}} - Character field \\Describes type of measurement associated with observation (either ``Estimate" or ``Standard Error")
	\item \orsVarName{\textbf{estimate\_type}} - Character field \\Further describes type of measurement associated with observation (our focus was on observations described as ``Percent")
	\item \orsVarName{\textbf{unit\_of\_measure}} - Character field \\Describes units of measurement associated with observation (our focus was on observations described as ``Percentage")
	\item \orsVarName{\textbf{additive\_group}} - Numeric field \\Exhibits one-to-one correspondence with \orsVarName{data\_element\_text} (i.e. requirement) \\Importantly, estimates of observations belonging to a given \orsVarName{occupation\_text} within a specific \orsVarName{additive\_group} should sum to 100\%; we will refer to these distinct combinations of occupation and requirement as \emph{\textbf{occupational groups}} moving forward
	\item \orsVarName{\textbf{value}} - Numeric field \\Measurement estimate (our focus was on percentage values describing mean distributions of requirements)
\end{enumerate}

In the filtered dataset, there were 420 different occupations, spanning 52 different requirements (yielding 21,840 occupational groups). Some of these requirements were manually constructed using similarity with other requirements and available data, the details of which can be found below (\sectionref{subsec:feat}). For our purposes, we eliminated the occupation \textit{All Workers}, since it was a summary measure of all occupations being considered in the ORS. This left us with 419 occupations and 52 requirements (for a total of 21,788 occupational groups). Of the 419 occupations, 22 were 2-digit SOC code categories which aggregate the values from the detailed occupations. These were treated identically to the specialised occupations in our analysis.





\subsection{Missing Data Problem} \label{subsec:missing_data}

There are two categories of missing information within the ORS data: (1) missing observations, and (2) missing estimates. There former were dealt with in a manner described in \sectionref{subsec:data_comp}, while the latter were the subject of imputation.

The known values represented in the data comprise 0.5343127 
of the potentially observable response population (this was computed based on the survey being executed to completion, described in \sectionref{subsec:data_comp}).
The missing amount is close to half of the total value (0.4656873),
but is spread over a larger number of observations (26,157
known estimates vs. 59,319
missing estimates), implying that the average missing value is 0.17 (compared to the average known value of 0.45). 
Since our model has prediction bounding for missing values, this is a great advantage because we narrow the potential for error with explicit model inputs.

\section{Background and Related Works} \label{sec:bg-and-related}

\subsection{Missingness} \label{subsec:missingness}

There are three main classes for missing data mechanisms \cite{rubin2004multiple}:

\begin{itemize}
\setlength\itemsep{0em}
\item Missing Completely at Random (MCAR) - no relationship between the missingness of the data and any values, observed or missing.
\item Missing at Random (MAR) - there is a systematic relationship between the propensity of missing values and the observed data, but not the missing data
\item Missing not at Random (MNAR) - there is a relationship between the propensity of a value to be missing and its values
\end{itemize}

MNAR is considered ``non-ignorable'' because the missing data mechanism itself has to be modeled as you deal with the missing data. The one mechanism that might drive missingness in the ORS is when the value is small, such that the cost of obtaining the value is high compared to its contribution to the overall survey. This would be a problem, but we believe to be of low impact as the values are bounded.
In contrast, MCAR and MAR are both considered `ignorable' because the model need not include any information about the missing data itself when generating imputations.

\subsection{Imputation Methodologies} \label{subsec:methods}

Multiple imputation (MI) is an effective method for handling missing values: it calculates a range of values as opposed to a single value in order to generate a confidence interval around the imputed value, which is better for making statements about a population overall. While there have been issues documented with some of the assumptions inherent in its application \cite{meng1994multiple}, progress in MI procedures and applications have shown it to be a robust method for handling missing data \cite{murray2018multiple}.


\subsubsection{MICE} \label{subsubsec:mice}

One implementation of MI is the Multiple Imputation by Chained Equations \textbf{(MICE)} method. This technique involves repeated imputations, treating the observed values as ``parameters'' that are connected via equations to produce a distribution of values rather than a single point estimate.

It is beyond the scope of this paper to detail all of the aspects of MICE, but the goals are achieved via derived mechanisms discussed below. We do note that MICE uses a series of regression models to estimate each missing value based on the other variables in the data. This allows for the modeling of each variable using the most appropriate distribution \cite{huang2020safe}.

We were unable to find an implementation of the MICE algorithm capable of dealing with the bounded nature of the ORS data, a feature that is a major advantage in the missing value inference. We settled on an approach using simulations for imputation, but since all of the data is bounded on [0,1], we presume they can be modeled by a single distribution type such as the $\beta$ distribution.


Briefly, each known estimate from the \orsVarName{value} field was simulated 10 times. Within an occupational group, negative correlation was enforced. The largest value in the occupational group was simulated first, and the opposite impact was proportionally spread across the other observations within the group. As a thought experiment to understand this, suppose an observed value in the data is 95 (0.95) and another is 2 (0.02). If we shock the 0.95 to be 0.98, it is clear that the value of the 0.02 observation must decrease to accommodate this. We note that other means for implementing correlation were researched, but proved to be unnecessarily complicated due to the bounded nature of potential values and the $\beta$ distribution. We discuss this procedure in greater detail in \sectionref{subsec:data_uncertainty}, where we address the issue of data uncertainty.

The end result of these simulations is a way to impute a distribution for each of the missing values, rather than a single point estimate. This provides a mean estimate and a standard deviation or error, which allows for calculating a confidence interval around the results.
We believe that these consistent simulations across points provide researchers a range of likely values that can be used to estimate model parameters in a manner similar to MICE.

\subsubsection{MICE Output vs. Our Model's Output} \label{subsubsec:mice-vs-model}

While MICE is a multiple input, single output model, our algorithm is a multiple input, multiple output model since the results are population values for different jobs and job requirements. We believe it is advantageous to model the outputs together due to the cross correlation between jobs and requirements. An early indicator of this came from initial exploratory data analysis, where we built a crude ensemble model to predict the values together using the original fields; this model performed well enough to set us off into the XGBoost approach. Another indicator is the good performance of k-fold validation based on the original observed data (see \sectionref{subsubsec:kfolds} for details).

\begin{figure}[!h]
    \centering
    \begin{measuredfigure}
    \includegraphics[height=5.35in,width=6.5in]{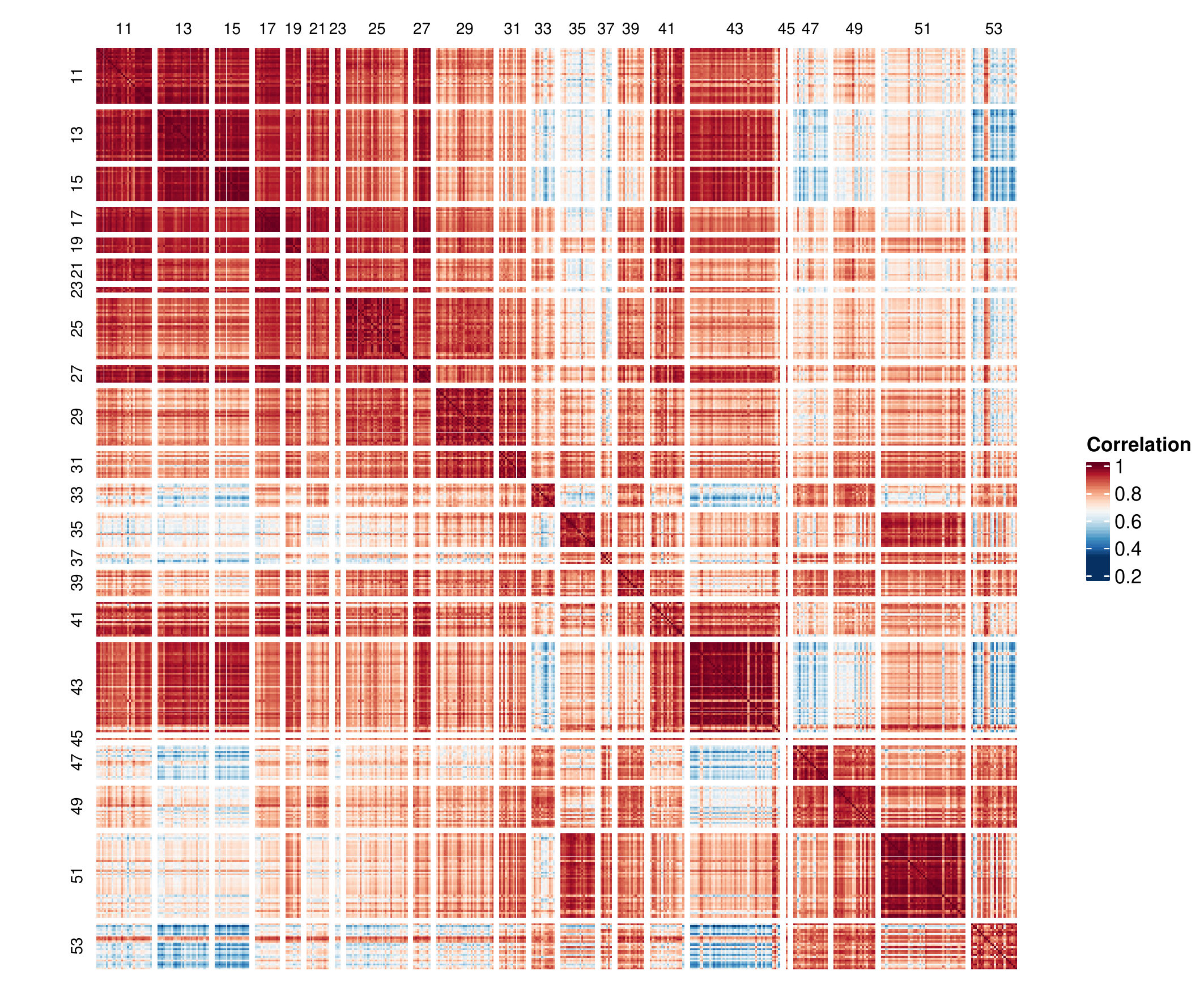}
    \end{measuredfigure}
    \caption{Heatmap of occupational correlation, organized by 2-digit SOC code. Descriptions of 2-digit SOC codes can be found at \cite{2digitSOC}. We note that intra-group correlation is high (major diagonal), and that similar 2-digit SOC groups (i.e., those comprised of similar occupations) exhibit higher inter-group correlation than dissimilar 2-digit SOC groups.}
    \label{fig:heatmap}
\end{figure}

Finally, at the risk of circular logic, we computed expected values for each occupational group by summing the frequency times the intensity times the population estimate which came from either the original data, or, if missing, the imputation. This provided a representation of each occupation in terms of expected values of each requirement. Details of this calculation are given in \sectionref{subsubsec:measuring-ELE}. These expected values were then used to calculate correlations between the occupations. The resulting heatmap is included below in \figref{fig:heatmap}, with the occupations sorted by 2-digit SOC codes (see \sectionref{subsubsec:soc_code} for a description of these codes). We can see there is significant correlation within SOC2 groups, and to a lesser extent within nearby SOC2 neighbors.

\subsubsection{Existing ORS Imputation Procedures} \label{subsec:existingORS}
We note that the BLS already has procedures in place to impute missing values within ORS survey responses \cite{righter2016imputation}. Their methods rely on a nearest neighbor approach that utilizes a systematic transfer of information from ``donors" (known estimates) to ``recipients" (missing estimates). While these existing methods do take into consideration details such as job proximity (based on SOC classification), requirement similarity, maximal retention of known data, and multi-layered imputation, they are limited in that they only use information from donors within strictly defined neighborhoods. In fact, the smart guessing procedure we use for initializing our models (discussed in \sectionref{subsec:smart_guess}) is similar to the existing imputation scheme used by the BLS. We add an ensemble machine learning (ML) approach to these procedures in the hopes of (a) utilizing more information already captured by the ORS in an empirically driven fashion, (b) measuring the error inherent in the imputations, (c) generating confidence intervals for our imputed estimates, and ultimately (d) developing an iterative approach to imputation that terminates based on errors measured throughout the prediction process.

\subsubsection{Ensemble Machine Learning Methods} \label{subsec:ensemble}

Ensemble methods are becoming very popular because they tend to be very powerful on problems out of the box with little tuning, and they are somewhat intuitive. The two most common tools are RandomForest and XGBoost. There is growing literature for using RandomForest for missing values, including for survey data \cite{stekhoven2012missforest}. The appeal is clear: many of the aspects of the MICE algorithm are included such as using multiple models, as the ensemble methods utilize many models by sub-sampling the data and predictors and then averaging the results.

\subsubsection{XGBoost} \label{subsec:xgboost}

We ultimately utilized the current implementation of XGBoost for key available features, namely:

\begin{itemize}
\setlength\itemsep{0em}
\item Bounds for observations
\item Weightings for observations
\end{itemize}

\paragraph{Bounds for Observations}\label{para:bounds}

XGBoost allows for the use of bounds on the data when fitting a model. The known estimates have standard errors associated with them that can be used to this end. Similarly, the amount of value that is left for the missing estimates is a known quantity (the sum of the known values within an occupational group subtracted from 1, call this $b_u$), so this can be used as an upper bound on the predictions for the missing estimates (the lower bound, $b_l$, is always 0). Thus, all missing values must fall in the interval $[0,b_u]$. However, this bounding does not completely eliminate the possibilities that (a) the estimates within an occupational group post-prediction would sum to a value not equal to 1, or (b) the individual predictions would fall outside of [0,1]. These concerns were addressed as an added step during the prediction process (described in detail in \sectionref{subsubsec:train-test}  and \sectionref{subsubsec:constraint_checking}).

\paragraph{Weightings for Observations}\label{para:weightings}

The ability to weight the influence of different observations on the fit process provided another advantage: lower weights could be assigned to guesses as compared to the known estimates. This was utilized in an iterative algorithm which started based on initial guesses for the missing values, derived from averaging the known residual amount over the number of missing values in an occupational group (this was called a ``naive guess"). In cases where there was more information for observations of other jobs within a 2- or 3-digit SOC group, a more nuanced approach was used to generate the initial guess (see \sectionref{subsec:smart_guess}). A full description of the weighting scheme that was ultimately used can be found in \sectionref{subsubsec:train-test}.


\section{Data Preprocessing} \label{sec:preproc}

Before applying models to generate predictions for missing values, the data had to be reformatted in a manner suitable to our analyses methods. This involved data completion (\sectionref{subsec:data_comp}), and feature selection and engineering (\sectionref{subsec:feat}).

\subsection{Data Completion} \label{subsec:data_comp}

Prior to predictive analysis, it was important to fill whatever missing information we could easily ascertain based on the structure of the data itself. This data completion step involved two processes, described below.

\subsubsection{Incomplete Occupational Groups} \label{subsubsec:incomplete_ogs}

One of the gaps in the data was the fact that some observations within an occupational group were unlisted (i.e., not only was the estimate missing, but the observation itself was not in the original data). To fix this issue, we went through the data and generated all possible levels within a given occupational group. We then merged these observations with those available in the original dataset to generate a ``master complete" dataset. Note, because these additional observations did not have estimates associated with them, they became part of the imputation problem moving forward.


\subsubsection{N-1 Group Completion}

Some of the observations with missing estimates (including some of those generated by the procedure described in \sectionref{subsubsec:incomplete_ogs}) could be filled in based on the available information. For example, some requirements were binary in nature, e.g. the requirement \textit{Prior work experience} has only two levels: YES, and NO. Given one estimate, the other could be calculated. Similarly, for requirements where all but one estimate was known, the final one could easily be determined. Each occupational group was assessed to determine whether it was such an ``N-1 group," and those that were had their missing estimate calculated and added to the data.

\subsection{Feature Selection and Engineering} \label{subsec:feat}

We settled on seven potential predictors for our analysis, listed below in \tabref{tab:predictors}, with ORS source field(s) and data type. Further details are provided below the table.

\begin{table}[h]
    \centering
    \begin{threeparttable}
    \begin{tabular}{@{}lrr@{}}
    	\toprule
        \textbf{Predictor} & \textbf{ORS source field(s)} & \textbf{Data type} \\
        \midrule
        Occupation & \orsVarName{occupation\_text} & Categorical\\
        Additive group & \orsVarName{additive\_group} & Categorical\\
        Frequency & \orsVarName{data\_element\_text}, \orsVarName{data\_type\_text} & Numeric\\
        Intensity & \orsVarName{data\_element\_text}, \orsVarName{data\_type\_text} & Numeric\\
        Requirement & \orsVarName{data\_element\_text} & Categorical\\
        SOC code & \orsVarName{upper\_soc\_code} & Categorical\\
        Requirement category & \orsVarName{data\_element\_text} & Categorical\\
        \bottomrule
    \end{tabular}
    \caption{Predictors and ORS source fields. Additive group was ultimately not used as a predictor in the imputation procedure.}
    \label{tab:predictors}
    \end{threeparttable}
\end{table}

\subsubsection{Occupation}
As noted previously, all occupations except for \textit{All Workers} were considered in this analysis.

\subsubsection{Synthetic Additive Groups}
Five additive groups were manually created in order to accurately group observations. The first four of these corresponded with the following levels of \orsVarName{data\_element\_text}: \textit{Pre-employment training: Certification}, \textit{Pre-employment training: Educational Certificate}, \textit{Pre-employment training: License}, and \textit{Literacy required}. They were assigned to additive groups 89, 91, 90, and 11, respectively. The final synthetic additive group, 78, corresponded to \textit{Sitting or standing/walking}. This group is further addressed under \sectionref{subsubsec:req}, below.

Additive group ultimately was not used as a predictor, as the same information was captured more effectively by the requirement predictor, described below in \sectionref{subsubsec:req}.

\subsubsection{Frequency} \label{subsubsec:frequency}
This predictor is a numeric transformation of the information in the \orsVarName{data\_type\_text} fields associated with each level of \orsVarName{data\_element\_text}. 
This predictor was manually generated,
and describes how often a requirement shows up in an occupation. An example of frequency assignments can be found below in \tabref{tab:lift_carry}. 
The full listing of frequency assignments can be found in \appref{app:A}.
When possible, values were assigned using guidance from the ORS Collection Manual \cite{ORSmanual}.


\subsubsection{Intensity} \label{subsubsec:intensity}
This predictor is a numeric transformation of the information in the \orsVarName{data\_type\_text} fields associated with each level of \orsVarName{data\_element\_text}. This predictor was manually generated, 
and describes the magnitude of a requirement. An example of intensity assignments can be found below in \tabref{tab:lift_carry}. 
The full listing of intensity assignments can be found in \appref{app:A}.
When possible, values were assigned using guidance from the ORS Collection Manual \cite{ORSmanual}.


\begin{table}[h]
    \centering
    \begin{threeparttable}
    \begin{adjustbox}{width=1\textwidth}
    \begin{tabular}{@{}ll|lrr@{}}
        \toprule
        \textbf{\orsVarName{data\_element\_text}} & \textbf{\orsVarName{data\_type\_text}} & \textbf{Requirement} & \textbf{Frequency} & \textbf{Intensity}\\
        \midrule
        Lifting/carrying Seldom & NONE & Lifting/carrying & 2 & 0 \\
        Lifting/carrying Seldom & NEGLIGIBLE & Lifting/carrying & 2 & 0.5 \\
        Lifting/carrying Seldom & $>$1 LBS, $<$=10 POUNDS & Lifting/carrying & 2 & 1 \\
        Lifting/carrying Seldom & $>$ 10 LBS, = 20 LBS & Lifting/carrying & 2 & 10 \\
        Lifting/carrying Seldom & $>$ 20 LBS, = 50 LBS & Lifting/carrying & 2 & 20 \\
        Lifting/carrying Seldom & $>$50 LBS, = 100 LBS & Lifting/carrying & 2 & 50 \\
        Lifting/carrying Seldom & $>$ 100 LBS & Lifting/carrying & 2 & 100 \\
        Lifting/carrying Occasionally &  NONE & Lifting/carrying & 33 & 0 \\
        Lifting/carrying Occasionally &  NEGLIGIBLE & Lifting/carrying & 33 & 0.5 \\
        Lifting/carrying Occasionally &  $>$1 LBS, $<$=10 POUNDS & Lifting/carrying & 33 & 1 \\
        Lifting/carrying Occasionally &  $>$ 10 LBS, = 20 LBS & Lifting/carrying & 33 & 10 \\
        Lifting/carrying Occasionally &  $>$ 20 LBS, = 50 LBS & Lifting/carrying & 33 & 20 \\
        Lifting/carrying Occasionally &  $>$50 LBS, = 100 LBS & Lifting/carrying & 33 & 50 \\
        Lifting/carrying Frequently &  NONE & Lifting/carrying & 67 & 0 \\
        Lifting/carrying Frequently &  NEGLIGIBLE & Lifting/carrying & 67 & 0.5 \\
        Lifting/carrying Frequently &  $>$1 LBS, $<$=10 POUNDS & Lifting/carrying & 67 & 1 \\
        Lifting/carrying Frequently &  $>$ 10 LBS, = 25 LBS & Lifting/carrying & 67 & 10 \\
        Lifting/carrying Frequently &  $>$ 25 LBS, = 50 LBS & Lifting/carrying & 67 & 25 \\
        Lifting/carrying Constantly &  NONE & Lifting/carrying & 100 & 0 \\
        Lifting/carrying Constantly &  NEGLIGIBLE & Lifting/carrying & 100 & 0.5 \\
        Lifting/carrying Constantly &  $>$1 LBS, $<$=10 POUNDS & Lifting/carrying & 100 & 1 \\
        Lifting/carrying Constantly &  $>$ 10 LBS, = 20 LBS & Lifting/carrying & 100 & 10 \\
        Lifting/carrying Constantly &  $>$ 20 LBS & Lifting/carrying & 100 & 20 \\
        \bottomrule
    \end{tabular}
    \end{adjustbox}
    \caption{Data transformation using \textit{Lifting/carrying} as an example. The two columns on the left (original data) were mapped to the three columns on the right (transformed data).}
    \label{tab:lift_carry}
    \end{threeparttable}
\end{table}

\subsubsection{Requirement} \label{subsubsec:req}
The \orsVarName{data\_element\_text} field was minimally reorganized (described below) and renamed as \textbf{requirement}. This field was used moving forward as a predictor in lieu of \orsVarName{data\_element\_text}.
Full details for each of these requirements can be found in \appref{app:A}.

\paragraph{Lifting/Carrying}
The original data contained four \orsVarName{data\_element\_text} levels related to \textit{Lifting/Carrying}-- \textit{Lifting/Carrying: Seldom}, \textit{Lifting/Carrying: Occasionally}, \textit{Lifting/Carrying: Frequently}, and \textit{Lifting/Carrying: Constantly}. These were reorganized under a single requirement, \textit{Lifting/Carrying}, mapping to frequencies 2, 33, 67, and 100, respectively (see \tabref{tab:lift_carry}).

\paragraph{Reaching}
The original data contained two \orsVarName{data\_element\_text} levels related to \textit{Reaching}-- \textit{Reaching overhead}, and \textit{Reaching at/below the shoulder}. These were reorganized under a single requirement, \textit{Reaching}, mapping to intensities 100 and 50, respectively.

\paragraph{Sitting or standing/walking}
The original data contained two levels of \orsVarName{data\_element\_text} related to \textit{Sitting or standing/walking}-- \textit{Sitting}, and \textit{Standing/walking}. These were reorganized under a single requirement, \textit{Sitting or standing/walking}, mapping to intensities 50 and 100, respectively. 

\subsubsection{SOC Code} \label{subsubsec:soc_code}
The field \orsVarName{upper\_soc\_code} contains fine-grained codes for classifying occupations based on the SOC system. This level of detail was not suitable for our analysis. However, SOC codes are structured such that the leading two digits indicate the broader industry, with the remaining digits specifying more detailed occupations within said industry. We peeled off the leading two digits, yielding 22 distinct codes, and used this as a predictor moving forward. The 2-digit SOC designations can be found online here \cite{2digitSOC}. We additionally used the leading three digits to generate a more fine-grained classification, yielding 95 distinct codes. This 3-digit SOC code was also used as a predictor moving forward.

\subsubsection{Requirement Category} \label{subsubsec:reqcat}
All of the occupational requirements assessed in the ORS fall into one of four categories, namely:
\begin{enumerate}
	\setlength\itemsep{0em}
	\item Cognitive and mental requirements (COG)
	\item Education, training, and experience requirements (EDU)
	\item Environmental conditions (ENV)
	\item Physical demands (PHY)
\end{enumerate}
	
These categories, and their constituent requirements, can be found online \cite{ORSmanual}. We note that the data used in this analysis (Wave 1) did not include any of the COG requirements.
The mapping of relevant requirements to categories can be found in \appref{app:A}.

\subsection{Advantages of Data Transformation}

\subsubsection{Dimensional Reduction} \label{subsubsec:dim_reduction}
Mapping the original data to \textbf{Frequency}, \textbf{Intensity}, and \textbf{Requirement} as described in \sectionref{subsubsec:frequency}-\ref{subsubsec:req} provided a reduction in  modeling dimensions from 204 to 50. 
A full description of this mapping and reduction can be found in \appref{app:A}, 
with a sample focusing on \textit{Lifting/carrying} in \tabref{tab:lift_carry}. Note that \orsVarName{data\_element\_text} and \orsVarName{data\_type\_text} (the first two columns in \tabref{tab:lift_carry}) are both categorical variables. There are 4 levels of \orsVarName{data\_element\_text}, and each of these levels corresponds to 5-7 different levels of \orsVarName{data\_type\_text}, resulting in 23 dimensions for the \textit{Lifting/carrying} data alone (i.e., each row in \tabref{tab:lift_carry} describes a dimension in the original data). Using the transformation described in \sectionref{subsubsec:frequency}-\ref{subsubsec:req}, this data is reduced to two dimensions, i.e. the two numerical variables Frequency and Intensity (the last two columns in \tabref{tab:lift_carry}).

Comparison of  model results from pre- and post- data transformations using an ensemble learning technique showed no loss in model performance. 

\subsubsection{Ordinal Value Advantage} \label{subsubsec:ordinal_advantage}
Another benefit of moving to a numeric representation of the relevant fields is improvement in modeling ability through the use of ordinal relations. For example, for the requirement \textit{Lifting/carrying}, we know that 100 lbs $>$ 50 lbs, but in the data's original form this relative/directional relation is lost to the model since these two values are considered two distinct, albeit related, classifications. The transformation we describe above preserves the original relationship between \orsVarName{data\_element\_text} and \orsVarName{data\_type\_text} while also adding numeric ordinal information, further allowing the model to infer predictions through the use of the relative values of the Intensity/Frequency fields.

There could be added value in terms of including directional information in the model. For example, someone who has 10 years of experience in a given occupation can perform a role that only requires 5 years of experience, but not the other way around. This was not included in the fit analysis, however it can be a key aspect for use in job mapping, discussed below in \sectionref{subsec:job-similarity}.

\section{Exploratory Data Analysis} \label{sec:eda}

\begin{figure}[h]
    \centering
    \begin{measuredfigure}
    \includegraphics[height=7.5cm,width=10cm]{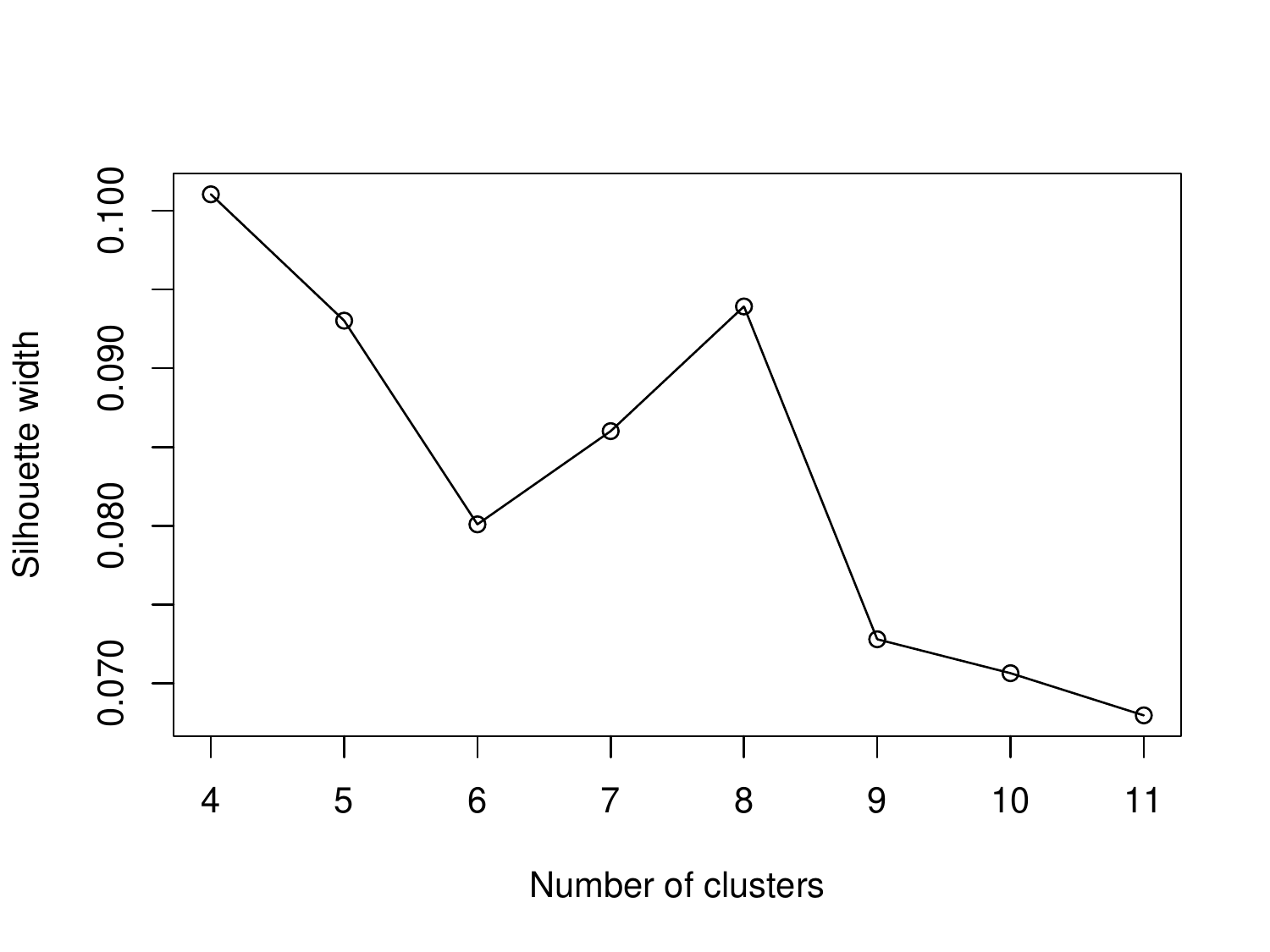}
    \end{measuredfigure}
    \caption{Silhouette plot. Analysis revealed a global maximum at 4 clusters, and a local maximum at 8. Upon inspection, these cluster assignments did not provide optimal visual clarity, so we chose to move forward with 5 clusters (which had the next largest silhouette width, seen above).}
    \label{fig:silwidth}
\end{figure}

Before moving forward with prediction, we sought to confirm that there was enough information within the existing (known) data to perform this imputation with a reasonable level of confidence. Having selected our predictors, we pursued exploratory analysis and visualization. The data was translated into a distance matrix using Gower's distance (on account of having mixed data types). Silhouette analysis was then performed to select the optimal number of clusters in the data, the results of which are shown and described in detail below in \figref{fig:silwidth}.

\begin{figure}[h]
    \centering
    \begin{measuredfigure}
    \includegraphics[height=10cm,width=15cm]{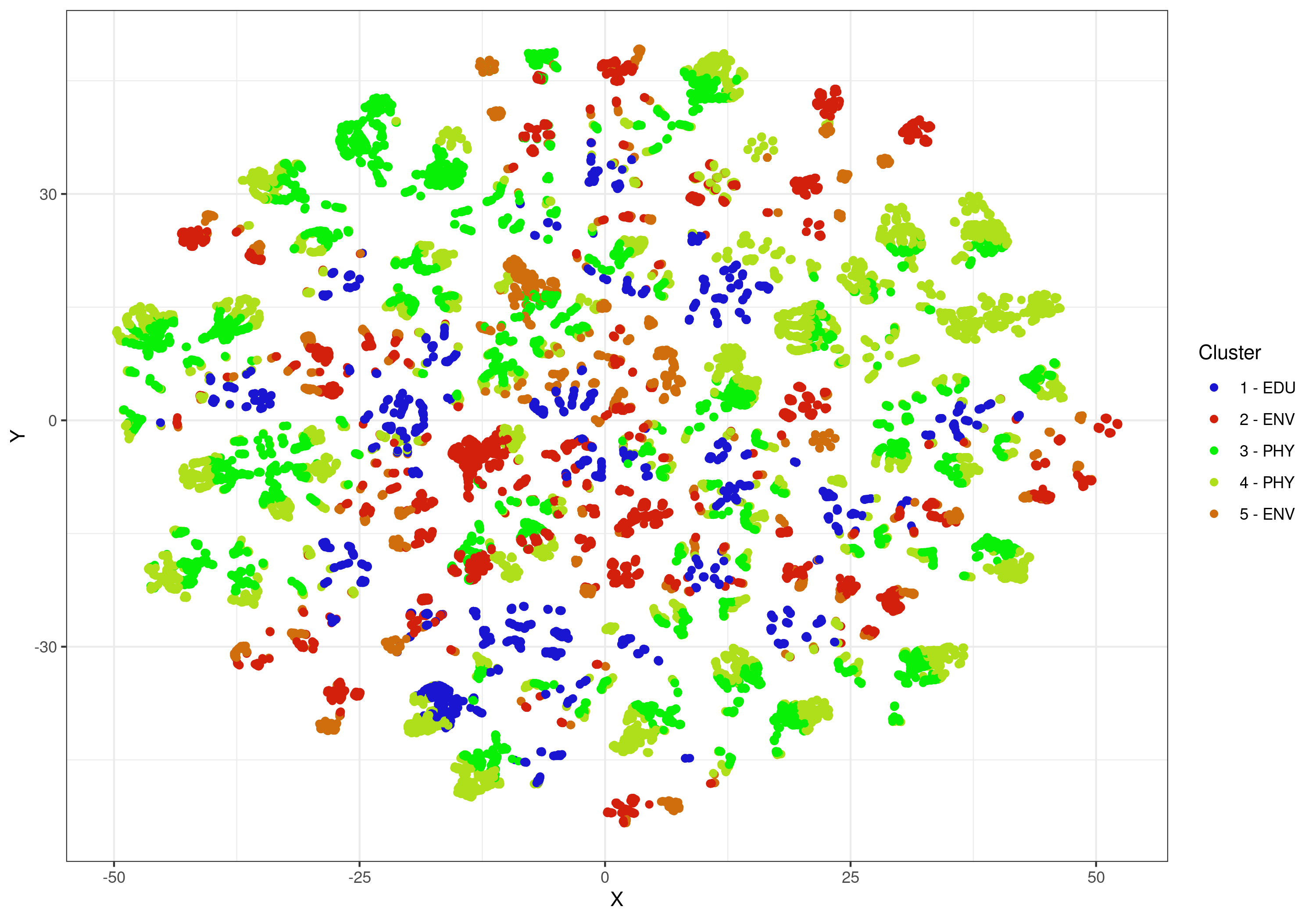}
    \end{measuredfigure}
    \caption{t-SNE plot. Clustering was primarily driven by the requirement category associated with each observation, as seen in \figref{fig:cluster_comp}, so we colored the points by their relevant factor and provided the associated requirement categories in the legend.}
    \label{fig:tsne}
\end{figure}

\begin{figure}[h]
    \centering
    \begin{measuredfigure}
    \includegraphics[height=5cm,width=10cm]{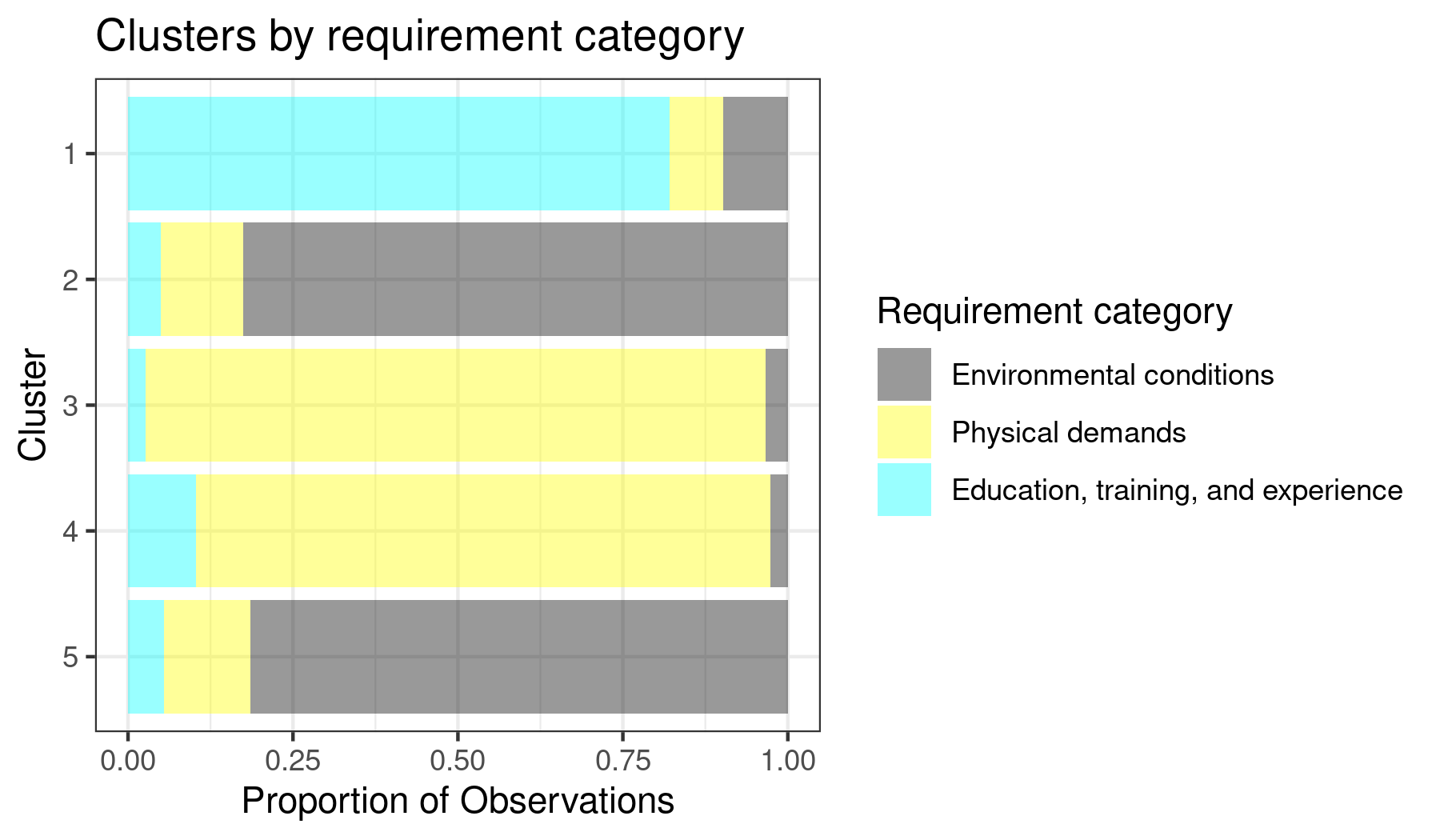}
    \end{measuredfigure}
    \caption{Cluster composition. We see the majority of the composition of each cluster is drawn from a single requirement category.}
    \label{fig:cluster_comp}
\end{figure}

The results of the silhouette analysis suggest an optimal cluster number of 5. We used the partitioning around medioids (PAM) method to generate our cluster assignments. Finally, the data was analyzed using the t-distributed stochastic neighbourhood embedding (t-SNE) method. The results of these analyses can be seen in \figref{fig:tsne} and \ref{fig:cluster_comp}.

In \figref{fig:tsne} we colored the points by their factor label and added the dominant requirement category. The latter was assigned after reviewing \figref{fig:cluster_comp}: each factor showed heavy influence from one of the three relevant requirement categories (EDU, ENV, or PHY). Within the clusters, there is some breakdown in \figref{fig:tsne} with regard to ``Presence" or ``Absence" of binary job requirements, from left to right. Based on this analysis we included requirement category as a predictor in our models.

\section{Sources of Uncertainty} \label{sec:uncertainty}

There are two key sources of uncertainty within our framework: \textbf{Model Uncertainty} due to the chosen models, and \textbf{Data Uncertainty} due to observed data uncertainty. We address these below.

\subsection{Model Uncertainty} \label{subsec:model_uncertainty}

Our choice of models has lower model uncertainty since they average many different models. While trying to capture this in our algorithm it was found to be close to zero for the observed data in the fit (discussed in further detail in \sectionref{subsubsec:measuring_model_uncertainty}). With this in mind, we focused on data uncertainty.

\subsection{Data Uncertainty} \label{subsec:data_uncertainty}

In order to address the potential impact of data uncertainty, simulated data was drawn from a $\beta$ distribution using, for all known observations, the \orsVarName{value} field ($\mu$) for the mean and the \orsVarName{std.error} field ($\sigma_e$) as the standard deviation. The standard deviation is a function of the number of observations and $\sigma_e$, but we do not know the number of observations. Trial and error of different choices to match $\sigma_e$ yielded problems as the upper bound value of the standard deviation, due to the bounded nature of the potential values, is a function of the mean value used in the $\beta$ distribution. This is described as follows:

$$ \sigma<=\mu*(1-\mu) $$

The maximum value $\sigma$ can take on is 0.25, which is when $\mu$ is equal to 0.5 . Note, as the value of $\mu$ approaches either 0 or 1, $\sigma$ approaches 0.
The variability in the data was significant, even using the lower value. If $\sigma_e$ was still too large to fit in the interval, the value was adjusted to be 95\% of the upper bound. This occurred 2,446 times while there were 10,187 values that were inside the limit.

With this in mind, we moved on to generating the simulations. Given some occupational group with observations $\{x_i\}$, recall that the sum of the $x_i$'s must be 1 and that there is structural negative correlation between them. We used a simple approach to reproduce this feature in our simulations by using the largest $x_i$ to generate the shocks and then scaling the resulting shock with negative correlation to the other smaller $x_i$'s. Formally, to generate a distribution for our imputations we generated shocked datasets assuming a $\beta$ distribution with lower bound of 0 and upper bound of 1, using the data's given standard error $\sigma_e$ as an approximation for the standard deviation for the largest observation, and offsetting the remaining observations within an occupational group with a proportional negative shock. Namely, if $\tilde{x}$ is the largest value of an occupational group $\{x_i\}$, then given the random shock $\epsilon$ to $\tilde{x}$ drawn from a $\beta$ distribution with mean $\tilde{x}$ and standard deviation $\sigma_e$, the remaining $x_i$'s are given by:

$$x_i=x_i-\frac{x_i}{\tilde{x}}*\epsilon$$

In this way the the population sum of 100\% is preserved and the standard deviation assumption is applied consistently across all observations within the occupational group. The negative correlation between the largest $x_i$ vs. the others is preserved.

Since we were only drawing 10 samples of $x_i$ from the $\beta$ distribution, we also ensured the mean $\tilde{x}$ and standard deviation $\sigma_e$ were reproduced in the small sample by subtracting the mean of the $x_i$'s, dividing by the standard deviation of the $x_i$'s, multiplying by $\sigma_e$, and then adding $\{x_i\}$. This ensured the shocks for each simulated observation were consistent with the values derived from the data across the small sample for each observation independently.

The newly generated datasets were then run through the imputation process and each missing value produced a range of predictions, generating both a mean estimate and a distribution thereof. In this way, missing values that were more sensitive to the initial assumptions show wider variation in their range of predictions, leading to broader confidence intervals for the imputed missing values. The details of this analysis are described in the following section, \sectionref{sec:imputation}.

\section{Imputation} \label{sec:imputation}

Imputation of the missing values involved a number of steps. These included implementing a more nuanced guessing procedure (\sectionref{subsec:smart_guess}), model tuning to establish optimal procedures and parameters based on known values (\sectionref{subsec:model_tuning}), and finally prediction of missing values, i.e. imputation (\sectionref{subsec:impute_missing}).

\subsection{Smart Guess Procedure} \label{subsec:smart_guess}

The first step of the imputation procedure was to generate an initial guess for missing values. These initial guesses (assigned lower weights) along with the known values (assigned maximal weight) would then be used in an iterative fashion to arrive at our estimates for the missing values. The naive approach for this initial guess would be to simply average the total remaining percentage within an occupational group over the number of missing values. For example, say for a given occupational group there were five total observations, with three of them having known estimates summing to 0.8. The total remaining available percentage would be $0.2$. Thus, the two missing estimates would each be assigned a naive guess of $0.2 / 2 = 0.1$

However, this approach ignores some of the structure built into the data. Namely, occupations within a given 2- or 3-digit SOC group are similar in nature, thus it is reasonable to assume that the requirements for the constituent occupations are also similar. With this in mind, we devised a way to use information from the other members of a given SOC group to produce an initial guess for a particular occupation.

First, within each SOC group, we searched each \orsVarName{additive\_group} (i.e., requirement) for the job with the ``best" distribution, i.e. the job with the maximum number of known values for a given requirement. If there were multiple best distributions, these were averaged and scaled to arrive at a single distribution. Then, each occupation within that \orsVarName{additive\_group} was assessed, and missing values filled in with an initial guess using the procedure outlined below.

For each occupation, the values associated with each of the levels of the relevant requirement were assigned to variables according to the presence (or absence) of data with respect to the best distribution. See \tabref{tab:smart-guess-vars} for an example, generalized for the \textit{Speaking} requirement.

\begin{table}[h]
    \centering
    \begin{threeparttable}
    \begin{tabular}{@{}lcccc@{}}
        \toprule
        \textbf{\textit{Speaking} level} & \multicolumn{2}{c}{\textbf{Data availability}} & \multicolumn{2}{c}{\textbf{Assigned variables}} \\
         & \textbf{Best distribution} & \textbf{Job$_a$} & \textbf{Best distribution} & \textbf{Job$_a$} \\
        \midrule
        NOT PRESENT & - & X & $m_{b1}$ & $k_{a1}$ \\
        SELDOM & - & - & $M_{b1}$ & $M_{a1}$ \\
        OCCASIONALLY & X & X & $K_{b1}$ & $K_{a1}$ \\
        FREQUENTLY & X & - & $k_{b1}$ & $m_{a1}$ \\
        CONSTANTLY & X & - & $k_{b2}$ & $m_{a2}$ \\
        \bottomrule
    \end{tabular}
    \caption{Variable assignment for our guessing scheme. In \textbf{Data availability}, ``-" indicates a missing value, and ``X" indicates a known value. In \textbf{Assigned variables}, the variables indicate whether the value is missing (m) or known (k) in one of the distributions (lower case), or both distributions (upper case).}
    \label{tab:smart-guess-vars}
    \end{threeparttable}
\end{table}

The variable assignments in \tabref{tab:smart-guess-vars} were then used to guess missing values as follows. A scaling factor $S$ was calculated as:

$$S = \frac{1-\sum_1^i{K_{ai}}}{1-\sum_1^i{K_{bi}}} $$

This scaling factor was multiplied by $\sum_1^i{k_{bi}}$ to arrive at an estimate for $\sum_1^i{m_{ai}}$, with adjustments for boundary violations (values outside of [0,1], sums greater than 1, etc.). Then, it follows that:

$$\sum_1^i{M_{ai}} = 1 - (\sum_1^i{K_{ai}} + \sum_1^i{k_{ai}} + \sum_1^i{m_{ai}})$$

To actually assign guesses to missing values, $\sum_1^i{m_{ai}}$ was distributed equally among all $m_{ai}$, and $\sum_1^i{M_{ai}}$ was distributed equally among all $M_{ai}$. Again, adjustments were made for boundary violations. In cases where Job$_a$ had no information, it was assigned whatever known values were in the best distribution, and the remaining observations were assigned the naive guess. In cases where Job$_a$ contained information, but there was no overlap with the best distribution, missing values were assigned the naive guess. This approach allowed us to maximally use the information that was already available to fill in missing data across similar occupations.

\subsection{Model Tuning} \label{subsec:model_tuning}

Model tuning was accomplished using the known estimates. However, in order to generate ``predictions," some of the known estimates were thrown out, and these observations were treated as ``missing" data. Note, these mock missing data were assigned lower weights in the model training process. The manner in which data were excluded/simulated was two fold. We first used a train-test approach to tune the model parameters for the final imputation step. We then used a k-folds approach to determine convergence iterations for our extended ensemble approach (more details in \sectionref{subsubsec:kfolds}). There were a number of steps in this processes, described in detail in \sectionref{subsubsec:train-test} and \sectionref{subsubsec:kfolds}.

\subsubsection{Train-Test Split for Parameterization}  \label{subsubsec:train-test}

We first sought to optimize the model parameters. These parameters were used for the k-folds stage of tuning, and then eventually for the final imputation step. The basic algorithm for the train-test approach is outlined below in \algref{alg:train-test}, with further details following. Note that this procedure used only the known data, and did not utilize the smart guessing procedure described in \sectionref{subsec:smart_guess}.

\begin{algorithm}[h]
    \SetAlgoLined
    Split $data$ into $train$ and $test$ sets\\
    $i=0$\\
	\While {r $\geq$ 0.001}{
		\eIf{$i==0$}{
			$data[test] =$ NA\\
			$known_{actual} = data[train]$\\
			\vspace{0.25cm}
			$prediction_i = $ guess($data$)\\
		}{
			$data[train] = known_{actual}$\\
			$data[test] = prediction_{i-1}[test]$\\
			\vspace{0.25cm}
			Set bounds for observations\\
			Set weights for observations\\
			Fit $model$ on $data$ using bounding and weights\\
			\vspace{0.25cm}
			$prediction_{i}$ = predict on $data$ using $model$\\
			$r$ = RMSE($prediction_{i-1}$,$actual$) - RMSE($prediction_{i}$,$actual$)\\
		}
		$i++$
    }
    $convergence\_iteration = i-2$
    \caption{Iterative train-test approach for model tuning.}
    \label{alg:train-test}
\end{algorithm}

\paragraph{Initialization} 
In this tuning step, 80\% of the data was used as the training set. Another 10\% of the data was kept aside as the test set, and this was held constant throughout the procedure. As a result, and unlike in the k-folds approach (see \sectionref{subsubsec:kfolds}), this meant that only 10\% of the observations were treated as ``missing." The final 10\% of the data was used as a validation set for parameter selection. The initial guess (iteration 0) for the mock missing values was a simple naive guess, based on the remaining available percentage in an occupational group and the number of missing observations (both true and mock). For example, say for a given occupational group there were five total observations, with three of them having known estimates summing to 0.8. Now suppose two of these known observations (with a total value of 0.4) were thrown out as mock missing. The total remaining available percentage would be $0.2 + 0.4 = 0.6$. Thus, the two mock missing observations would each be assigned a ``guess" value of $0.6 / 4 = 0.15$.

\paragraph{Iteration, Constraint Checking, and Convergence}
After the initial iteration, the values for the mock missing data were updated with the previous iteration's predictions. We initially tried using a weighted average of the previous iteration's input data and output prediction (e.g., $d_i = a*d_{i-1} + b*p_{i-1}$, where $d$ denotes input data and $p$ the output predictions). We tried multiple pairings of $(a,b)$, namely $(0.5,0.5)$, $(0.25,0.75)$, and $(1.0,0.0)$. We settled on using $(1.0,0.0)$, as it made more intuitive sense to let the predictions fully guide subsequent iterations of the model. However, the model output had to be adjusted to meet known constraints on the data. Specifically, predictions had to be adjusted to account for (a) those that were $<0$ or $>1$, and (b) sums of occupational groups that were $\neq 1$. The former were addressed by setting these values to 0 and 1, respectively, and the latter were addressed by scaling the guesses by the remaining available percentage. Building on the initialization example,
, if the predictions of the two mock missing estimates were returned as 0.2 and 0.3, they would be scaled to 0.08 and 0.12, respectively. Note that these \textit{adjusted} predictions became the input data for the next iteration. Also note that the model predicts values for the known observations, but these observations were fixed to their actual value during training. This iterative approach was repeated until the model converged, i.e. the difference in root mean square error (RMSE) between consecutive predictions was $< 0.001$. We note that \cite{stekhoven2012missforest} uses a similar approach, relying on RMSE, to define model convergence.

\paragraph{Weighting Guesses} XGBoost allows differential weighting of data that is used to train the model. Because the mock missing data were populated with guesses, it was appropriate to leverage this ability. We tried a number of approaches, using different initial weights for the guessed values as well as different functions by which the weights increased over multiple iterations. 
We eventually settled on a starting weight of 0.25 for naive guesses (mock missing data), increasing linearly by 0.05 with each iteration to a maximum of 0.75.
Known values were assigned a fixed weight of 1.


\paragraph{Results} Using the iteration scheme, constraint checking, convergence criteria, and weighting scheme described above, the train-test procedure was completed for multiple model parameterizations. Specifically, we altered $nrounds$, $max\_depth$, and $eta$. After 
6 different combinations of values for these parameters, we settled on the following: $nrounds=200$, $max\_depth=14$, and $eta=0.6$.
This specific combination was selected based on minimizing the mean absolute error (MAE) of predictions from the validation set.
This parameterization was then used in both the k-folds tuning stage (\sectionref{subsubsec:kfolds}), and the final imputation step (\sectionref{subsec:impute_missing}).

\subsubsection{K-Folds} \label{subsubsec:kfolds}

In the next stage of tuning, we used a k-folds approach (k=10) to arrive at convergence iterations for our extended ensemble approach (further details below). The basic algorithm is outlined below in \algref{alg:kfolds}, with more detailed explanation of the specific steps following. Note that this procedure used all of the data (known and missing values).

\begin{algorithm}[h]
	\SetAlgoLined
	\SetKw{KwBy}{by}
    $fold$ = split $data$ into $k$ folds\\
    $i=0$\\
	\While {r $\geq$ 0.001}{
		\eIf{$i==0$}{
			\For{$f = 1$ \KwTo $k$ \KwBy $1$}{
			    $train = fold[-f]$\\
			    $test = fold[f]$\\
			    \vspace{0.25cm}
			    $data[test] =$ NA\\
			    $known_{actual} = data[train]$\\
			    \vspace{0.25cm}
			    $prediction_i = $ smartGuess($data$)\\
			}
		}{
			\For{$f = 1$ \KwTo $k$ \KwBy $1$}{
			    $train = fold[-f]$\\
			    $test = fold[f]$\\
			    \vspace{0.25cm}
				$data[train] = known_{actual}$\\
				$data[test,missing] = prediction_{i-1}[test,missing]$\\
				\vspace{0.25cm}
				Set bounds for observations\\
				Set weights for observations\\
				Fit $model$ on $data$ using bounding and weights\\
				\vspace{0.25cm}
				$prediction_{i}$ = predict on $data$ using $model$\\
			}
			$r$ = RMSE($prediction_{i-1}$,$actual$) - RMSE($prediction_{i}$,$actual$)\\
		}
		$i++$
    }
    $convergence\_iteration = i-2$
    \caption{Iterative k-folds approach for model tuning.}
    \label{alg:kfolds}
\end{algorithm}

\paragraph{Initialization}
The known data were initially split into 10 equal folds, which were held constant throughout the rest of the procedure. Each fold was treated as the test set once, and the estimates in the test set were thrown out and treated as ``missing" data to simulate imputation. In this way, after running through all the folds, all observations were treated as ``missing" once. These mock missing values were replaced with an initial guess that was calculated using the smart guess procedure outlined in \sectionref{subsec:smart_guess}. The actual missing values were then appended to the known data, initialized with a smart guess in the same manner as the mock missing data. The model was trained on the full dataset (known and all guessed values), and then applied back to the data to generate new predictions for the test set (as well as the actual missing values). 

\paragraph{Subsequent Iterations and Convergence} The iterative guessing, constraint checking, and convergence criteria were identical to those used in the train-test approach. The weighting scheme was slightly different, however, because this tuning approach used the smart guessing procedure to produce the initial predictions. Naive guesses received a starting weight of 0, and smart guesses received a starting weight of 0.5. The weight of naive guesses was fixed at 0, but that of smart guesses was increased linearly by 0.05 with each iteration, to a maximum of 0.75. Known values were assigned a fixed weight of 1. This scheme was also used in the final imputation step described below (\sectionref{subsec:impute_missing}).

\begin{figure}
    \centering
    \begin{measuredfigure}
    \includegraphics[height=5cm,width=15cm]{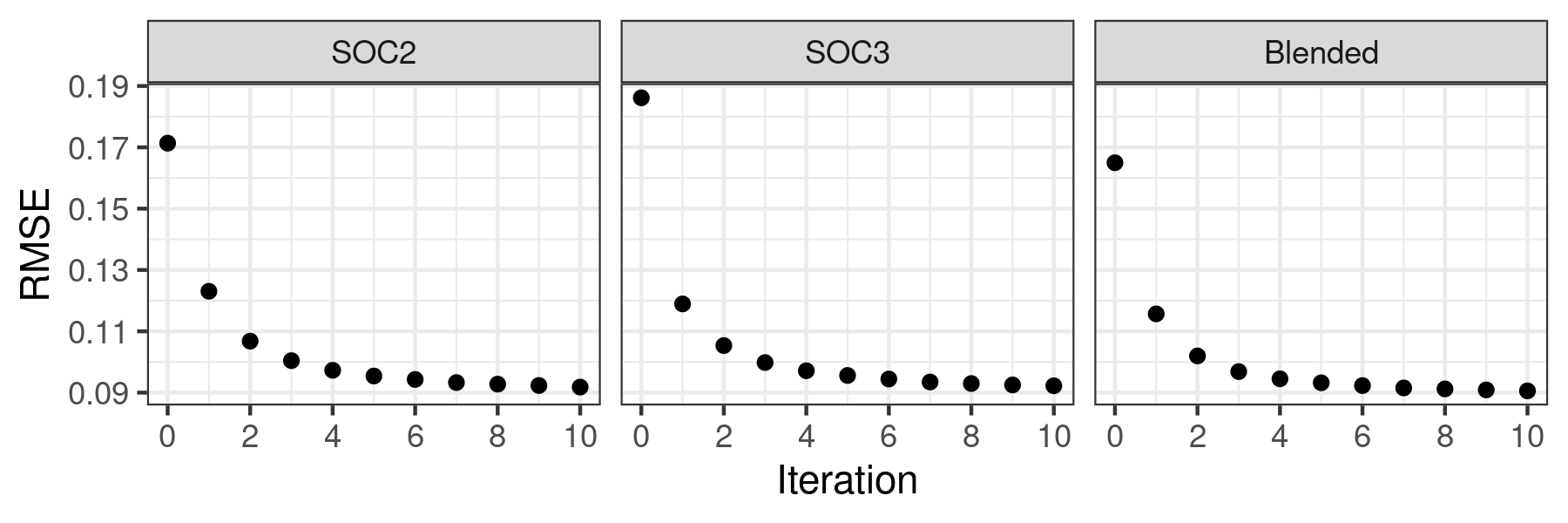}
    \end{measuredfigure}
    \caption{RMSE of the two KFCV models, along with the final blended model, by iteration. Iteration 0 corresponds to the initial smart guess, with each subsequent iteration corresponding to (adjusted) model predictions.}
    \label{fig:rmse}
\end{figure}

\begin{figure}[h]
    \centering
    \begin{measuredfigure}
    \includegraphics[height=8cm,width=12.8cm]{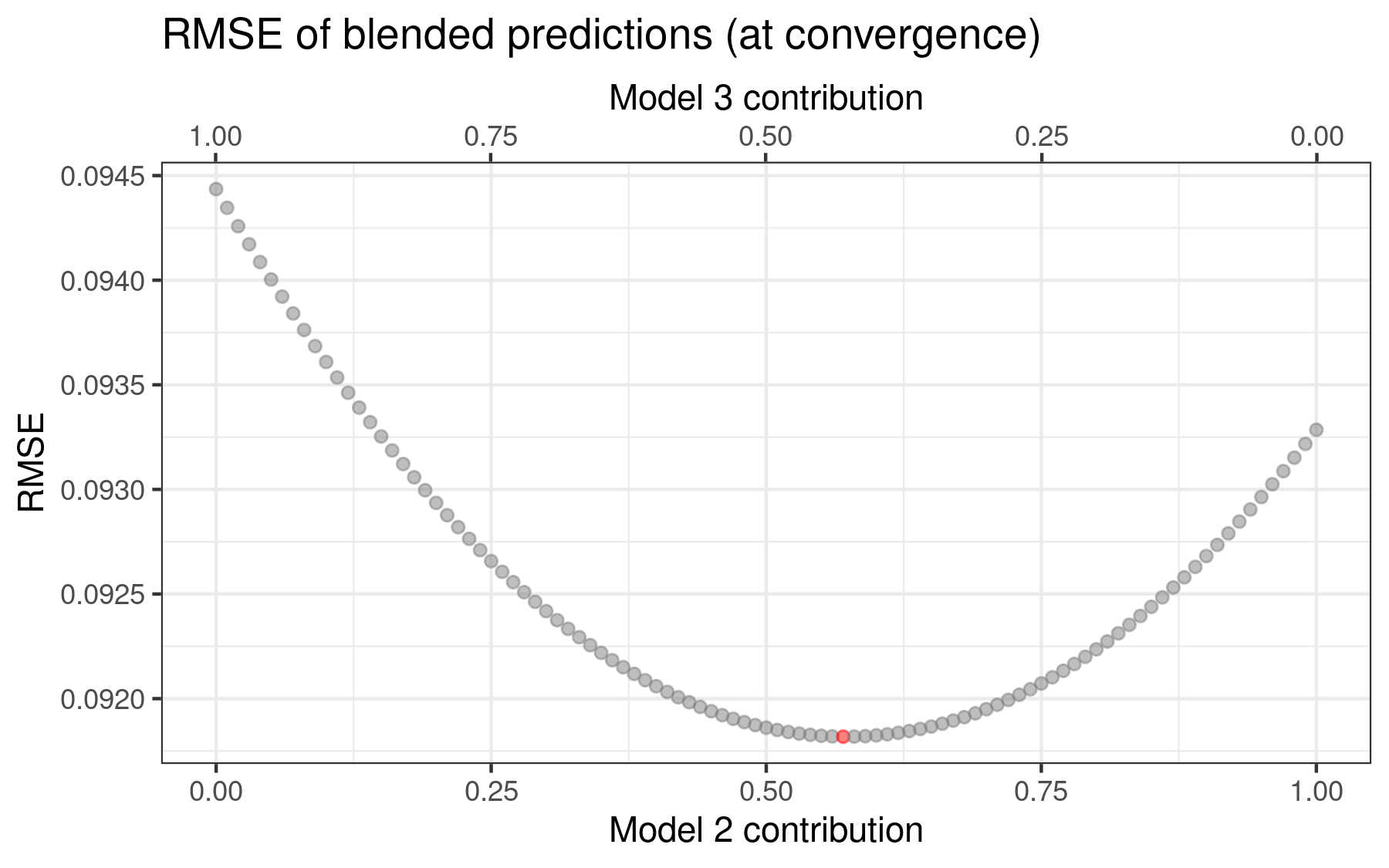}
    \end{measuredfigure}
    \caption{Calculation of model blending ratio from KFCV results. The predictions at the convergence iteration of each of the constituent models were combined in various ratios, and the RMSE of the resulting blended prediction was calculated. The ratio that minimized this RMSE was 57(SOC2):43(SOC3), represented by the red point on the plot.}
    \label{fig:blending_models}
\end{figure}

\paragraph{Extended Ensemble Approach} The analysis described above was completed using both 2-digit SOC groups for the smart guess procedure, as well as 3-digit SOC groups in parallel. The convergence iteration was identified for both analyses (iteration 7 for the SOC2 model, and iteration 6 for the SOC3 model). The predictions from the two convergence iterations were blended in a ratio derived from minimizing the RMSE. From this analysis we arrived at a blending ratio of 57:43 (SOC2:SOC3). See \figref{fig:rmse} and \tabref{tab:rmse} for the RMSE of predictions from each of the two analysis streams, as well as from the blended model, by iteration. \figref{fig:blending_models} illustrates the blending ratio analysis. We report on this blended result in the following section. Importantly, this ensemble approach was also used throughout the remaining prediction step, namely \sectionref{subsec:impute_missing}.

\begin{table}
    \centering
    \begin{threeparttable}
    \begin{tabular}{@{}lrrr@{}}
        \toprule
        \textbf{} & \multicolumn{3}{c}{\textbf{RMSE}} \\
        \midrule
        \textbf{} & \textbf{SOC2 model} & \textbf{SOC3 model}  & \textbf{Blended model} \\
        \midrule
        Iteration 0 (smart guess) & 0.17135482 & 0.18616357 & 0.16497943\\
        Iteration 1 & 0.12305518 & 0.11893894 & 0.11569240\\
        Iteration 2 & 0.10678853 & 0.10535381 & 0.10197901\\
        Iteration 3 & 0.10043104 & 0.09981186 & 0.09686256\\
        ... & ... & ... & ...\\
        Convergence & 0.09328480 & 0.09443532 & 0.09181852\\
        \bottomrule
    \end{tabular}
    \caption{Tabular representation of data displayed in \figref{fig:rmse}, through convergence. We can see that as the iterations progress, the blended model consistently has a lower RMSE than either of the two constituent models from which it is derived.}
    \label{tab:rmse}
    \end{threeparttable}
\end{table}

\begin{figure}
    \centering
    \begin{measuredfigure}
    \includegraphics[height=15cm,width=14cm]{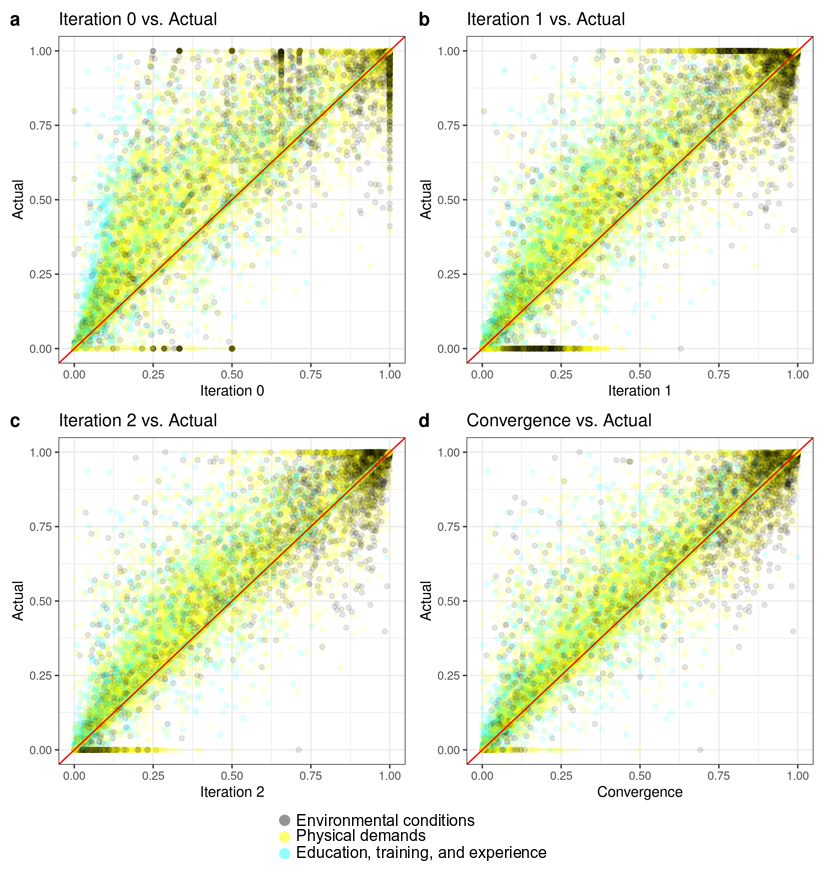}
    \end{measuredfigure}
    \caption{KFCV predicted estimates vs. their actual values by iteration. Shown are iterations 0 (output of smart guess procedure) through 2, as well as the results at termination (the final prediction). We can see a convergence of the predictions toward their actual values.}
    \label{fig:kfolds_plot}
\end{figure}

\paragraph{Results}
A visualization of the iterative process can be seen in \figref{fig:kfolds_plot}. 
The actual value versus prediction is plotted to show the progression from the initial smart guess to the (adjusted) prediction for the first two iterations of the model and at convergence.
The first set of guesses are crude because many are just simple uniform spreading between the missing values, and the various vertical lines can be mapped to whether an occupational group has two missing values, three missing values, all the way up to eight missing values. 

It can be seen that the XGBoost is moving the predictions to the actual values as the predictions are swept toward the diagonal at each step, most dramatically after the first epoch. The upper right hand corner represents high population concentrations approaching 100 percent, so we see the XGBoost effectively sweeps the initial evenly spread predictions to the actual values effectively at each step. We note that the blue points, representing the EDU requirement category which contains the requirement with the highest number of possible observations (\textit{Minimum education level}, with 9 levels), have the furthest to move and have the highest error.

We have also included in \tabref{tab:kfolds_errors} the mean error (ME), standard deviation of errors (SD$_e$), and $R^2$ of the predictions vs. actual values for each iteration until convergence. Note that these were computed using the data used to produce \figref{fig:rmse}, \figref{fig:blending_models}, \figref{fig:kfolds_plot}, and \tabref{tab:rmse}, i.e. by combining each of the 10 test folds per iteration to generate a complete dataset of mock missing values (all based on known observations) for each iteration.

\begin{table}[h]
    \centering
    \begin{threeparttable}
    \begin{tabular}{@{}lrrr@{}}
        \toprule
        \textbf{} & \textbf{ME} & \textbf{SD$_e$}  & \textbf{$R^2$} \\
        \midrule
        Iteration 0 (smart guess) & 0.04491778 & 0.15875006 & 0.8297732\\
        Iteration 1 & 0.03316830 & 0.11083803 & 0.9161088\\
        Iteration 2 & 0.02994266 & 0.09748602 & 0.9351158\\
        Iteration 3 & 0.02808683 & 0.09270285 & 0.9413804\\
        ... & ... & ... & ...\\
        Convergence & 0.02515772 & 0.08830646 & 0.9468794\\
        \bottomrule
    \end{tabular}
    \caption{Error measures for iterated k-folds validation, comparing predictions vs. actual values of the blended model. Note that ME and SD$_e$ decrease from one iteration to the next, and $R^2$ increases.}
    \label{tab:kfolds_errors}
    \end{threeparttable}
\end{table}

We believe this sets a reasonable upper bound on what the errors would be on new data predictions using all of the data due to (a) a larger quantity of known data, and (b) greater clarity around the missing values since the distribution for each occupational group is better specified (i.e., the potential for error decreases even faster due the bounding at 1 for the total population). Importantly, the convergence iterations and blending ratio derived from this k-folds approach were used in the final imputation step described in \sectionref{subsec:impute_missing}.


\subsection{Impute Missing Data} \label{subsec:impute_missing}

Having tuned our model, we moved on to imputing the missing values using the same iterative, ensemble methods described previously. Our approach relied on training the model on both the known estimates, and the guessed missing estimates. The basic algorithm for doing so is outlined in \algref{alg:impute}, with further details following.

\begin{algorithm}[h]
    \SetAlgoLined
    $known_{actual} = data[known]$\\
    $i=0$\\
	\While {$i \leq C$}{
		\eIf{$i==0$}{
		    $prediction_i$ = smartGuess($data$)\\
		}{
			$data[known] = known_{actual}$\\
			$data[missing] = prediction_{i-1}[missing]$\\
		    \vspace{0.25cm}
		    Set bounds for observations\\
		    Set weights for observations\\
		    Fit $model$ on $data$ using bounding and weights\\
		    \vspace{0.25cm}
		    $prediction_{i}$ = predict on $data$ using $model$\\
		}
		$i++$
    }
    \caption{Iterative imputation procedure, completed for each of 10 simulations. Note that $C$ is the convergence iteration determined by the k-folds validation procedure (see \algref{alg:kfolds}).}
    \label{alg:impute}
\end{algorithm}

\subsubsection{Initialization} \label{subsubsec:init}
Recall that the known values were simulated 10 times, as per the procedure outlined in \sectionref{subsec:data_uncertainty}. The result of this was 10 distinct data sets (simulations) to perform imputation on. For each simulation, the data was first subjected to the smart guessing procedure. These guesses were used as a starting point for the model iteration process. 

\subsubsection{Iteration, Constraint Checking, and Convergence} \label{subsubsec:constraint_checking}

The full data set (known and guessed estimates) was fed into an XGBoost fit, and the resulting model was used to predict the missing values. Some the of the resulting predictions were $<0$ or $>1$. These predictions were adjusted to 0 and 1, respectively (as in \sectionref{subsubsec:train-test}). The estimates (both known and predicted) within occupational groups were then checked to sum to a total value of 1; if they did not, the predicted values only (not the known observations) were adjusted in proportion to their prediction value to scale the sum to 1, as described in \sectionref{subsubsec:train-test}. The assumption here is that the model's relative prediction value gives a probability weighting of the the value. These were then joined with the known observations and again a fit was produced, and the missing values were predicted and adjusted. This procedure was repeated until the model had converged (at iteration 7 for the SOC2 model and iteration 6 for the SOC3 model, as per the results in \sectionref{subsubsec:kfolds}). 



\subsubsection{Results and Confidence Intervals} \label{subsubsec:impute_results}

The end result of running \algref{alg:impute} over each of the simulations was a distribution of 10 predictions per missing value. This allowed for us to calculate a mean prediction and its relevant 95\% confidence interval for each of the missing values, a result which is more useful than a single point estimate. In \tabref{tab:CIs} we provide a sample of these calculated confidence intervals (and associated mean predictions). Of the total 59,319 missing estimates, we were unable to calculate confidence intervals for 832 of them. Of these, 811 observations had a predicted value of 0 across all 10 simulations (one such observations is included in \tabref{tab:CIs}), and the remaining 21 observations had identical, non-zero predictions across all 10 simulations.


\begin{sidewaystable}
    \footnotesize
    \centering
    \begin{threeparttable}
    \begin{tabular}{@{}llllr@{}}
        \toprule
        \textbf{SOC2 code: Occupation} & \textbf{Additive group: Level} & \textbf{Lower Bound} & \textbf{Upper Bound} & \textbf{Mean prediction} \\
        \midrule
        11: Construction Managers & 36: SELDOM & 0.107757349 & 0.156272543 & 0.132014946\\
        11: Architectural and Engineering Managers & 25: NONE & 0.349589101 & 0.400369893 & 0.374979497\\
        11: Advertising and Promotions Managers & 25: $>50$ LBS, = 100 LBS & 0.003605466 & 0.020919323 & 0.012262395\\
        11: General and Operations Managers & 24: $>10$ LBS, = 20 LBS & 0.026148978 & 0.053150248 & 0.039649613\\
        13: Tax Preparers & 68: MEDIUM WORK & 0.012660474 & 0.047278176 & 0.029969325\\
        13: Fundraisers & 41: OCCASIONALLY & 0.063864607 & 0.095040466 & 0.079452536\\
        15: Software Developers, Applications & 24: NONE & 0.067670815 & 0.083341524 & 0.07550617\\
        15: Information Technology Project Managers & 25: $>50$ LBS, = 100 LBS & 0.003516028 & 0.011832868 & 0.007674448\\
        19: Environmental Scientists and Specialists, Including Health & 73: NONE & 0.020942507 & 0.05607407 & 0.038508289\\
        25: Vocational Education Teachers, Postsecondary & 10: SHORT DEMO ONLY & 0.010109042 & 0.01774264 & 0.013925841\\
        27: Editors & 32: OCCASIONALLY & 0.06465778 & 0.113281536 & 0.088969658\\
        29: Pharmacy Technicians & 42: FREQUENTLY & 0.008508049 & 0.015172574 & 0.011840311\\
        29: Licensed Practical and Licensed Vocational Nurses & 51: SELDOM & 0.000919444 & 0.002100302 & 0.001509873\\
        29: Opticians, Dispensing & 51: NOT PRESENT & 0.961028119 & 0.970955258 & 0.965991688\\
        29: Emergency Medical Technicians and Paramedics & 44: CONSTANTLY & 0.003287085 & 0.01603633 & 0.009661707\\
        29: Acute Care Nurses & 16: NOT PRESENT & 0.07370114 & 0.097411484 & 0.085556312\\
        31: Phlebotomists & 36: SELDOM & 0.089535678 & 0.120575394 & 0.105055536\\
        35: Food Preparation Workers & 10: OVER 10 YEARS & 0.000480806 & 0.003713051 & 0.002096928\\
        37: Building and Grounds Cleaning and Maintenance Occupations & 47: CONSTANTLY & 0.004624571 & 0.009285582 & 0.006955076\\
        39: Childcare Workers & 10: $>1$ MONTH,  = 3 MONTHS & 0.042169977 & 0.048106535 & 0.045138256\\
        43: Eligibility Interviewers, Government Programs & 73: HIGH SCHOOL VOCATIONAL & 0.000506788 & 0.005435561 & 0.002971175\\
        51: Cabinetmakers and Bench Carpenters & 53: NO & 0.716297997 & 0.780385328 & 0.748341662\\
        51: Stationary Engineers and Boiler Operators & 47: FREQUENTLY & 0.009095747 & 0.019708674 & 0.014402211\\
        53: Automotive and Watercraft Service Attendants & 29: FREQUENTLY & 0.02913031 & 0.040746434 & 0.034938372\\
        53: Laborers and Freight, Stock, and Material Movers, Hand & 73: ASSOCIATE'S VOCATIONAL & NaN & NaN & 0\\
        \bottomrule
    \end{tabular}
    \caption{Sample 95\% confidence intervals, calculated from predictions of missing estimates across 10 simulations.}
    \label{tab:CIs}
    \end{threeparttable}
\end{sidewaystable}


\subsubsection{Measuring Model Uncertainty} \label{subsubsec:measuring_model_uncertainty}
Earlier we mentioned that one of the sources of uncertainty within our framework is the models themselves (see \sectionref{subsec:model_uncertainty}). We note that during imputation, not only were the missing values predicted, but also the known ones. With each iteration, the known observations were reset to their actual values prior to predicting in the subsequent iteration, but the predictions were used to try and roughly quantify the uncertainty contributed by the models. To do so, we calculated both the mean absolute error (MAE) and the mean error (ME) between the actual values of known observations and the predictions generated by the model. This was done for each of the 10 simulations. At convergence, the average (mean) MAE across all 10 simulation was 0.01967782, and the average ME was 0.004064048. We conclude from this analysis that the model uncertainty is $\pm 0.0197$, and believe this to be a tolerable level of error.

\section{Summary} \label{sec:summary}

In this paper we have demonstrated a method by which to impute values for missing estimates within the Occupational Requirements Survey administered by the U.S. Bureau of Labor Statistics. Our method leverages many of the inherent features of the data to arrive at a distribution of predictions. These features include the bounded nature of the data (all observations must be on the interval $[0,1]$), negative correlation within an occupational group (percentages that must sum to 100\%), quantified errors of known observations, and similarity between different occupations within a 2- or 3-digit SOC group.

We first used the known errors to generate 10 simulations of the known estimates in order to generate a distribution of predictions, which allowed us to ultimately compute confidence levels for our imputations rather than single point estimates. We then utilized an iterative approach using XGBoost to generate our predictions for each simulation, iterating the models until they reached convergence. We then used the predictions from the convergence iteration of each simulation to calculate a mean prediction and its 95\% confidence interval.

In the next section (\sectionref{sec:applications}), we discuss possible applications of the dataset with the imputed values, as well as a generalized application of our method.

\section{Applications and Future Work} \label{sec:applications}

Imputing values for the missing observations within the ORS broadens the scope of what can be accomplished using this data. We propose two applications:

\begin{enumerate}
	\setlength\itemsep{0em}
	\item Computing job similarity on the basis of occupational requirements. This information could be leveraged to identify a wider range of potential placements for disability beneficiaries in jobs that match their current and/or expected abilities (\sectionref{subsec:job-similarity}).
	\item Updating and improving survey methodology. Doing so would facilitate streamlining and refining the ORS to emphasize capturing the most meaningful data (\sectionref{subsec:survey-methods}).
\end{enumerate}

In addition, we believe that the architecture of our imputation methodology can be generalized and applied to other survey data that conforms to certain specifications (\sectionref{subsec:generalizability}).

\subsection{Job Similarity} \label{subsec:job-similarity}

One of our proposed applications of the (imputed) ORS data is computing the similarity between different occupations. We describe two separate methods for doing so, detailed in \sectionref{subsubsec:measuring-overlap} and \sectionref{subsubsec:measuring-ELE}.

\subsubsection{Measuring Similarity using Population Distribution Overlap} \label{subsubsec:measuring-overlap}

If we have a full distribution of population percentages for two jobs, we can measure the overlap of these occupations by taking the product of their weightings to yield a value on the interval $[0,1]$. Further analysis of the requirement and the overlap can help lead to understanding of jobs that are similar or dissimilar.

If $\omega_{1ir}$ is the population mean of the $i^{th}$ level of the $r^{th}$ requirement for Job 1 (average of simulation predictions for missing values, and actual value for known observations), and $\omega_{2ir}$ is the same for Job 2, then we say that the overlap of $r^{th}$ requirement ($O_r$) for these two jobs is:

$$O_r = \sum_{i}^{} \omega_{1ir}*\omega_{2ir}$$

\begin{figure}[h]
    \centering
    \begin{measuredfigure}
    \includegraphics[height=6cm,width=15cm]{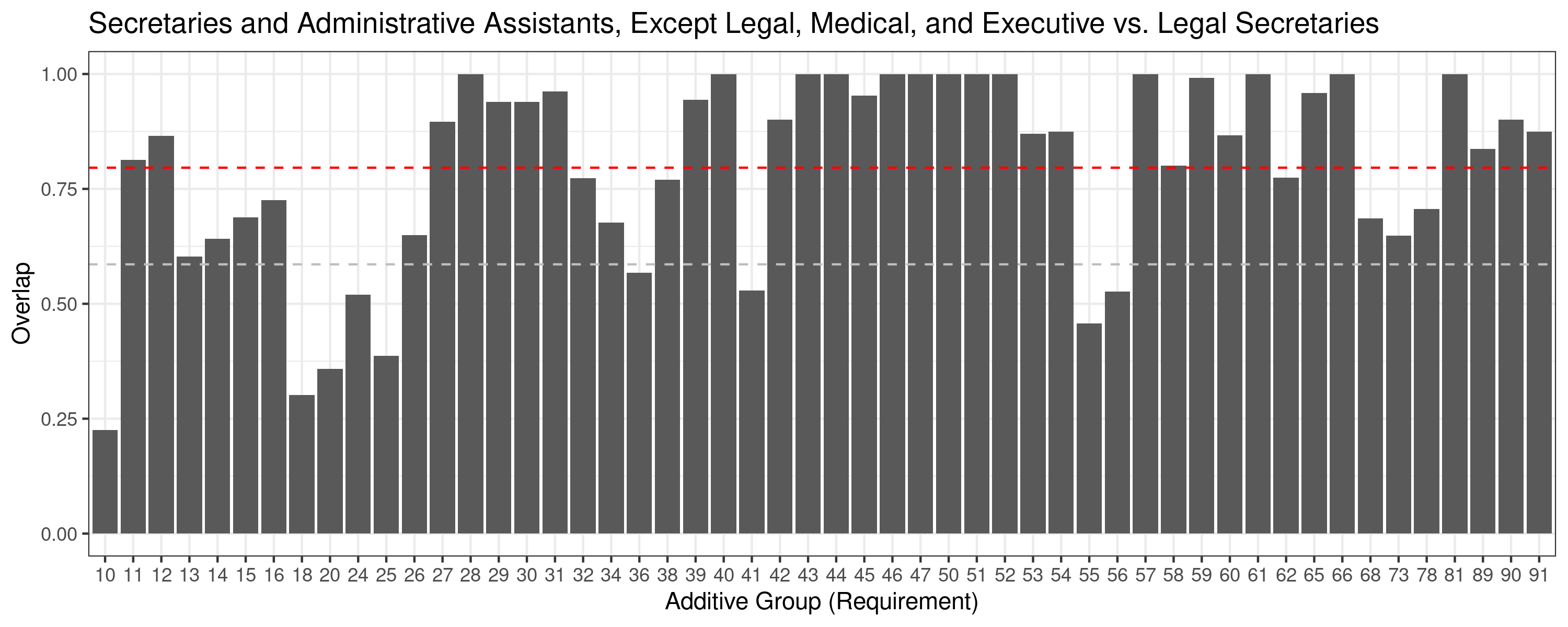}
    \end{measuredfigure}
    \caption{Overlap of General and Legal Secretaries. Mean +/- SD (dashed lines) assumes equal weighting of captured activity.}
    \label{fig:legSec-v-genSec}
\end{figure}

\begin{figure}[h]
    \centering
    \begin{measuredfigure}
    \includegraphics[height=6cm,width=15cm]{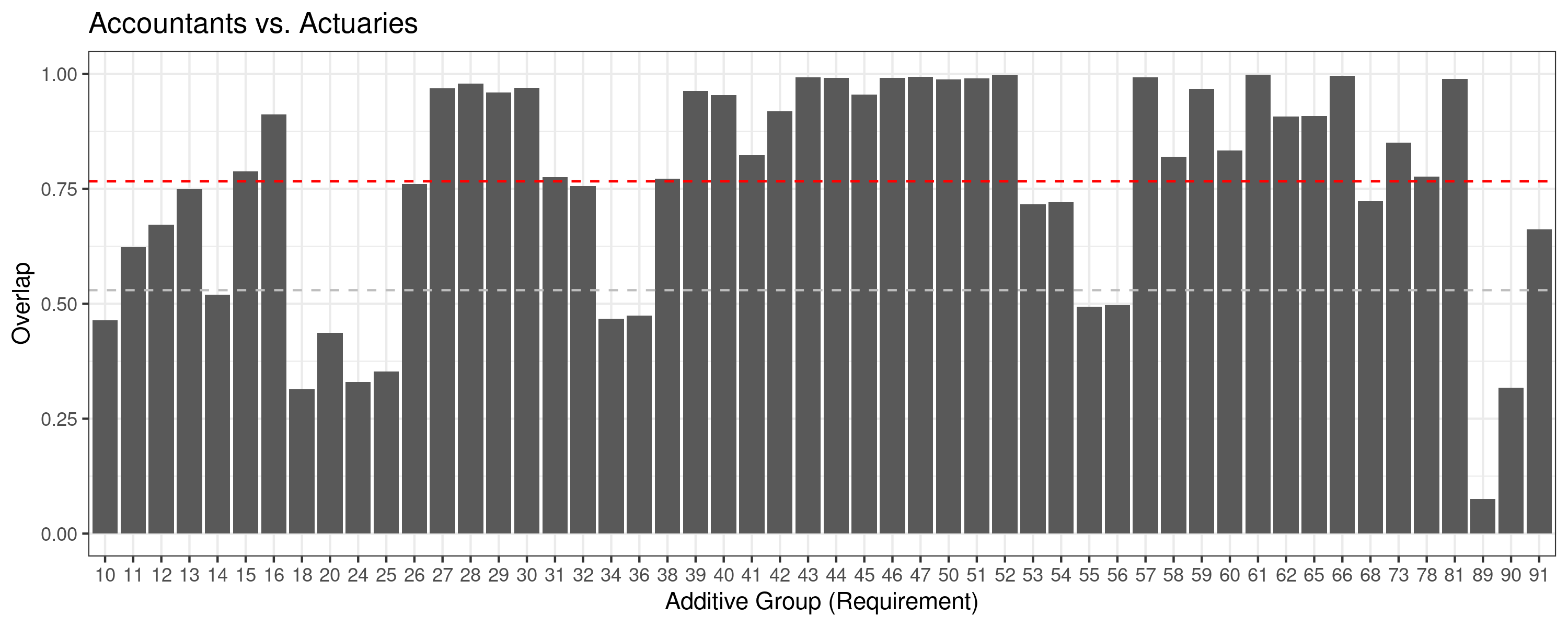}
    \end{measuredfigure}
    \caption{Overlap of Accountants and Actuaries. Mean +/- SD (dashed lines) assumes equal weighting of captured activity.}
    \label{fig:acct-v-act}
\end{figure}

\begin{figure}[h]
    \centering
    \begin{measuredfigure}
    \includegraphics[height=6cm,width=15cm]{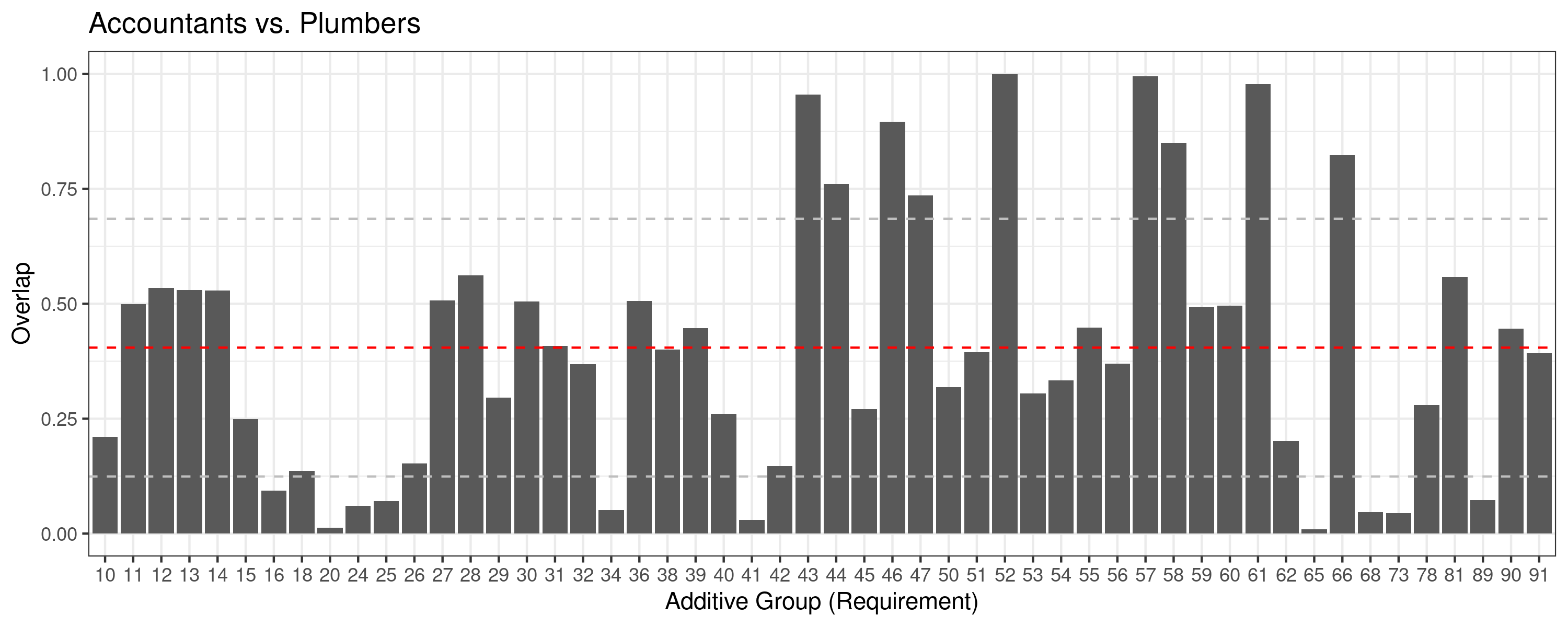}
    \end{measuredfigure}
    \caption{Overlap of Accountants and Plumbers. Mean +/- SD (dashed lines) assumes equal weighting of captured activity.}
    \label{fig:acct-v-plumb}
\end{figure}

Below we visualize the overlap between three pairs of occupations: \textit{Legal Secretaries} vs. \textit{General Secretaries} (\figref{fig:legSec-v-genSec}), \textit{Accountants} vs. \textit{Actuaries} (\figref{fig:acct-v-act}), and \textit{Accountants} vs. \textit{Plumbers} (\figref{fig:acct-v-plumb}). The first draws a comparison between two occupations we would expect to be very similar, the second does so for two occupations we expect are somewhat similar, and the third does so for two seemingly disparate occupations. These anticipated differences can be seen, to an extent, in the average overlap values (dashed red lines in the plots). 

\subsubsection{Measuring Similarity using Expected Level of Effort} \label{subsubsec:measuring-ELE}

Another measure we developed for comparing occupations across requirements was what we termed ``Expected Level of Effort” (ELE). This is a weighted average of the frequency and intensity times the population estimate for the various requirements. A low frequency/low intensity/low population estimate would result in a lower level of effort, and the converse for high.

For each occupational group, we calculated ELE as an expected value $E$ of frequency times intensity as follows, where $\mu_j$ is the mean population prediction across all 10 simulations for the $j^{th}$ observation, and $F_j$ and $I_j$ are the frequency and intensity of the $j^{th}$ observation, respectively:

$$E=\sum_j{\mu_j*F_j*I_j}$$

By calculating this for every occupation, we obtain a level of effort for each requirement which we can then use in a correlation calculation across occupations (see \sectionref{subsubsec:mice-vs-model}). For an application of ELE to improving survey design, see \sectionref{subsubsec:id-low-var-req}.

\subsubsection{Necessary but Redundant Requirements} \label{subsubsec:necessary-but-redundant}

There are a number of requirements that are common between jobs that do not provide differentially useful information. Take for example the requirement of \textit{Keyboarding: Traditional} (\orsVarName{additive\_group} 20). When considered on an equal basis to all other requirements, it implies that the nearest occupation for any job with this requirement is a typist. For a more useful aggregate metric, the various requirements should be weighted in some fashion prior to calculating a mean overlap value (dashed red lines on \figref{fig:legSec-v-genSec} - \figref{fig:acct-v-plumb}).

\subsubsection{Directional Information} \label{subsubsec:directional-information}

Much of the data has a directional aspect. For example, if job requires carrying 100 pounds then anyone who qualifies at that level can also do the job below that intensity threshold. This does not work the other way around, as there is no assurance that workers in a job that requires carrying 50 pounds will also be able to carry 100 pounds.

Another directional aspect is qualification. For example, \textit{Accountants} and \textit{Actuaries} both require education at a similar level, but it is unlikely that an accountant can use their background to do actuarial work, nor the other way around. So while the education levels may overlap, the actual qualifications likely do not.
On the other hand, a general management job class may also have similar education requirements to \textit{Actuaries} and \textit{Accountants}, and either of these two professions could slot into such a management position, but someone in the management role is highly unlikely to be qualified for either accounting or actuarial practice.

A robust overlap analysis would identify and account for these directional, non-transitive, and directionally-transitive properties.

\subsection{Survey Methodology Changes} \label{subsec:survey-methods}

\subsubsection{Field Consistency} \label{subsubsec:field-consistency}
As evident from \sectionref{subsec:feat}, the  roles of the data fields \orsVarName{data\_element\_text} and \orsVarName{data\_type\_text} are somewhat mixed and not strictly defined. It would be useful to identify how/when the role of these fields cross over, and then update them to maintain consistent roles across the survey.

Additionally, the \orsVarName{additive\_group}s associated with \textit{Lifting/Carrying} are split into a number of groups that might be better described in redesigned system of data fields. Moreover, the weights in \orsVarName{data\_type\_text} have different ranges that are very close but do not match across the different \orsVarName{additive\_group}s associated with \textit{Lifting/Carrying}. Aligning the weights so the ranges are identical across all \orsVarName{additive\_group}s associated with \textit{Lifting/Carrying} would make for a more consistent approach to data collection.

Since much of the crossover for \orsVarName{data\_element\_text} and \orsVarName{data\_type\_text} takes place within \textit{Lifting/Carrying}, it should be addressed at the same time as the weight consistency issue.


\subsubsection{Identifying Low Variability Requirements} \label{subsubsec:id-low-var-req}


As discussed above in \sectionref{subsubsec:measuring-ELE}, we can use the sum of the product of the population estimate times the frequency times the intensity to calculate what we call Expected Level of Effort (ELE). This measure was used to generate the heatmap in \sectionref{subsubsec:mice-vs-model}. Another application would be to look at the variability of ELE within a requirement to help identify where the sources of differentiation lie within the survey. Recall, for each occupational group, we calculated ELE as:

$$E=\sum_j{\mu_j*F_j*I_j}$$

These values were then standardized by requirement. This was calculated as follows, where $E$ is the expected value of a given occupational group within a requirement, and $\mu_{E}$ and $\sigma_{E}$ are the mean expected value and standard deviation of the expected values for the same requirement, respectively:

$$ \frac{E_{} - \mu_{E}}{\sigma_{E}} $$

\begin{figure}[h]
    \centering
    \begin{measuredfigure}
    \includegraphics[height=6cm,width=15cm]{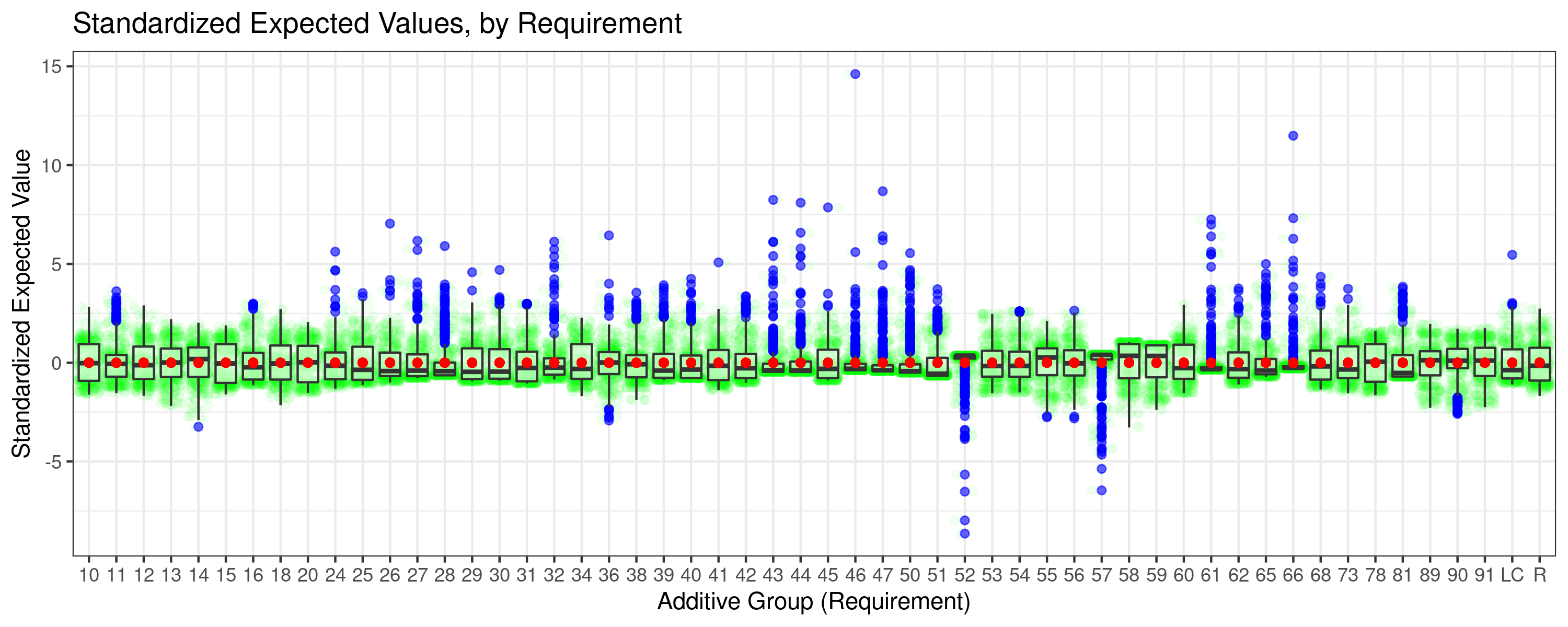}
    \end{measuredfigure}
    \caption{Boxplot of ELEs, standardized by \orsVarName{additive\_group} (requirement). Red points indicate standardized $\mu_{E}$'s (i.e., 0), and blue points indicate outliers. Requirements where $\mu_{E}$ and the median do not align are indicative of skew in the expected values. On the far right we have aggregated all the additive groups corresponding to the requirement \textit{Lifting/Carrying} (24, 25, 26, and 27) into a single distribution labeled LC. The same was done for \textit{Reaching} (additive groups 16 and 18), labeled R.}
    \label{fig:boxplot}
\end{figure}

A plot of these standardized values can be seen in \figref{fig:boxplot}. Low variability in a requirement (i.e., less spread in the expected values within a requirement) implies limited information content in that requirement (e.g. \orsVarName{additive\_group} 46 - requirement \textit{Humidity}). In contrast, higher variability in a requirement will likely do a better job of differentiating between the different occupations (e.g. \orsVarName{additive\_group} 73 - requirement \textit{Minimum education level}). This might be used by survey designers to increase the power of the survey elements, help them prioritize the components, or modify the collection process.

\begin{figure}[h]
    \centering
    \begin{measuredfigure}
    \includegraphics[height=5cm,width=12cm]{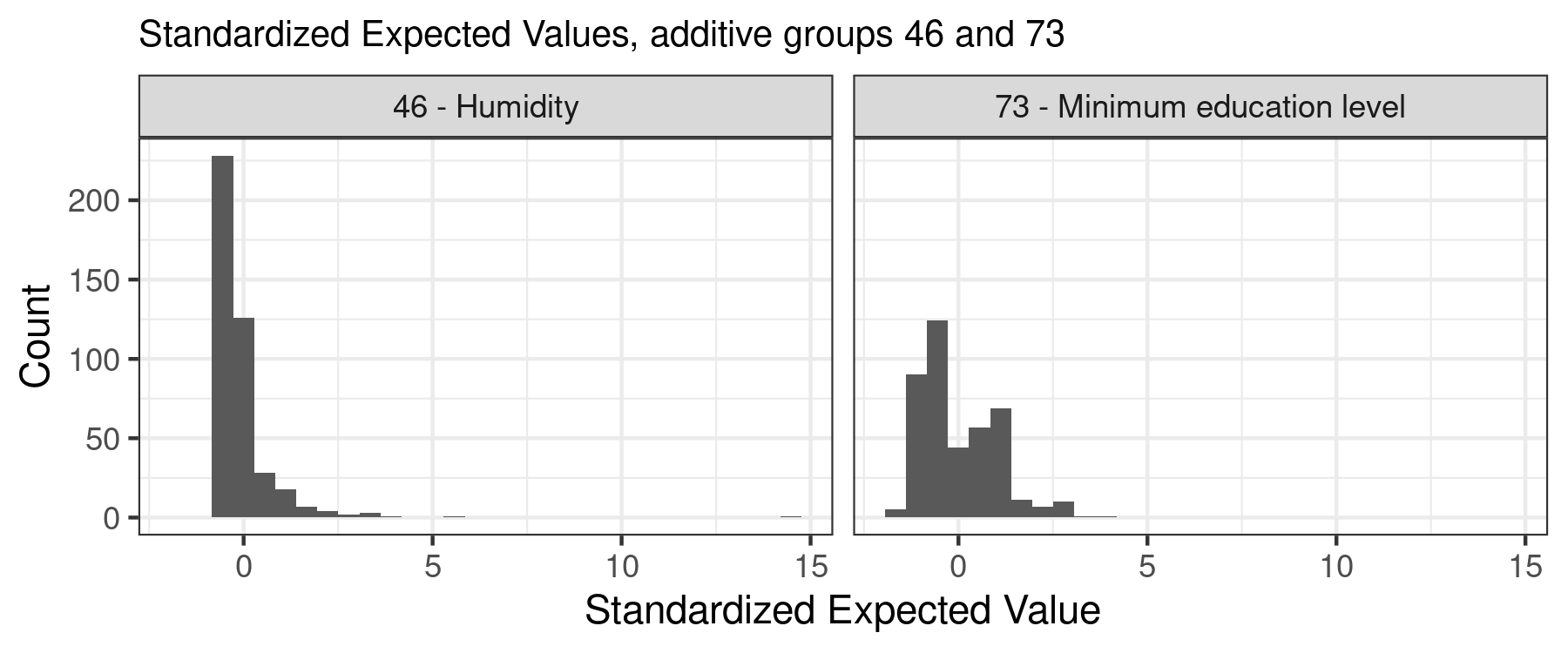}
    \end{measuredfigure}
    \caption{Distribution of standardized ELEs for additive groups 46 (\textit{Humidity}) and 73 (\textit{Minimum education level}).}
    \label{fig:adg46-73}
\end{figure}

For example, in \figref{fig:adg46-73}, we see that \textit{Minimum education level} has a spread of standardized ELE values around 0, whereas \textit{Humidity} is concentrated at 0 with a handful of outliers. This concentration suggests that only a small number of the returned results have any meaningful information, while the dispersion for \textit{Minimum education level} shows significant differentiation of the respondents. This is due to the fact that \textit{Humidity} does not differ greatly between various jobs: it rarely occurs as an occupational requirement (most respondents report that \textit{Humidity} is NOT PRESENT). Survey designers might consider reframing the question in such a way that it elicits a more nuanced collection of responses. Survey end users may consider if there are areas where greater differentiation would help in their analyses (this would also improve the performance of the imputation procedures presented in this paper).

\begin{figure}[h]
    \centering
    \begin{measuredfigure}
    \includegraphics[height=6cm,width=15cm]{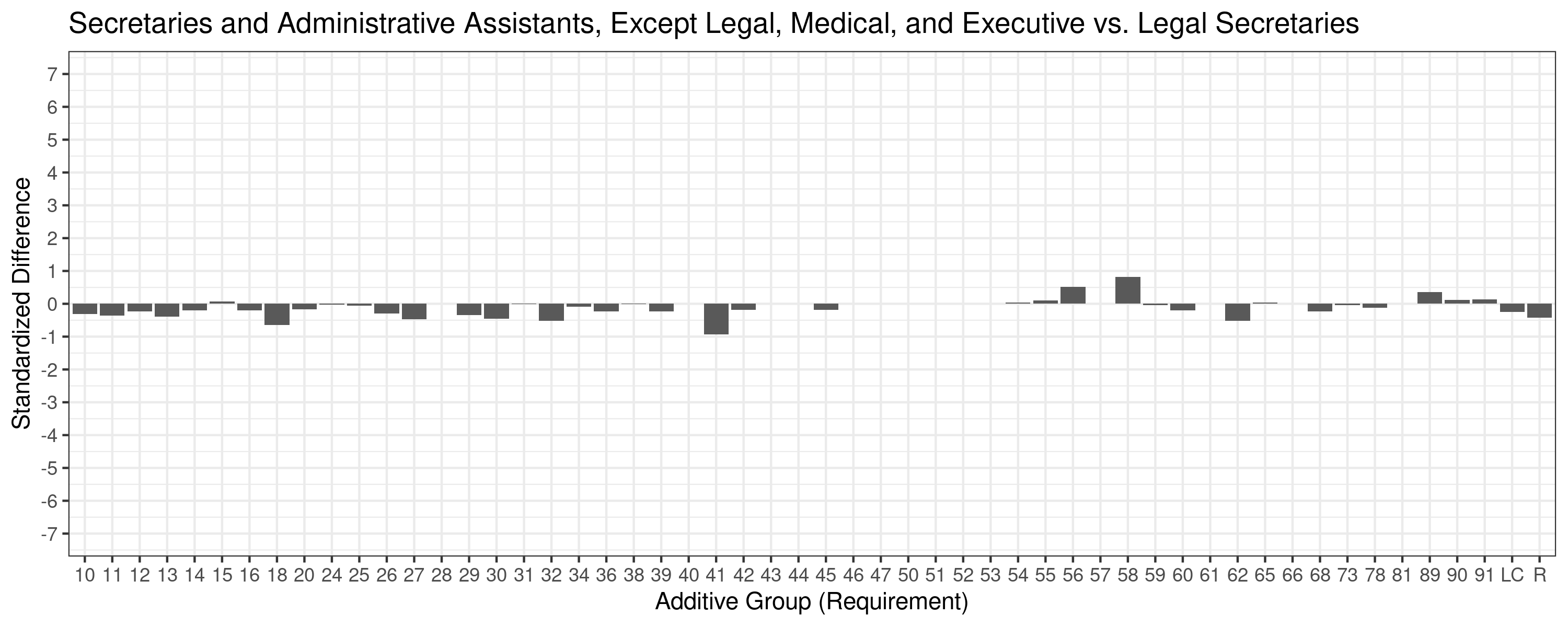}
    \end{measuredfigure}
    \caption{Difference in standardized ELEs between General Secretaries and Legal Secretaries.}
    \label{fig:diff_legSec-v-genSec}
\end{figure}

\begin{figure}[h]
    \centering
    \begin{measuredfigure}
    \includegraphics[height=6cm,width=15cm]{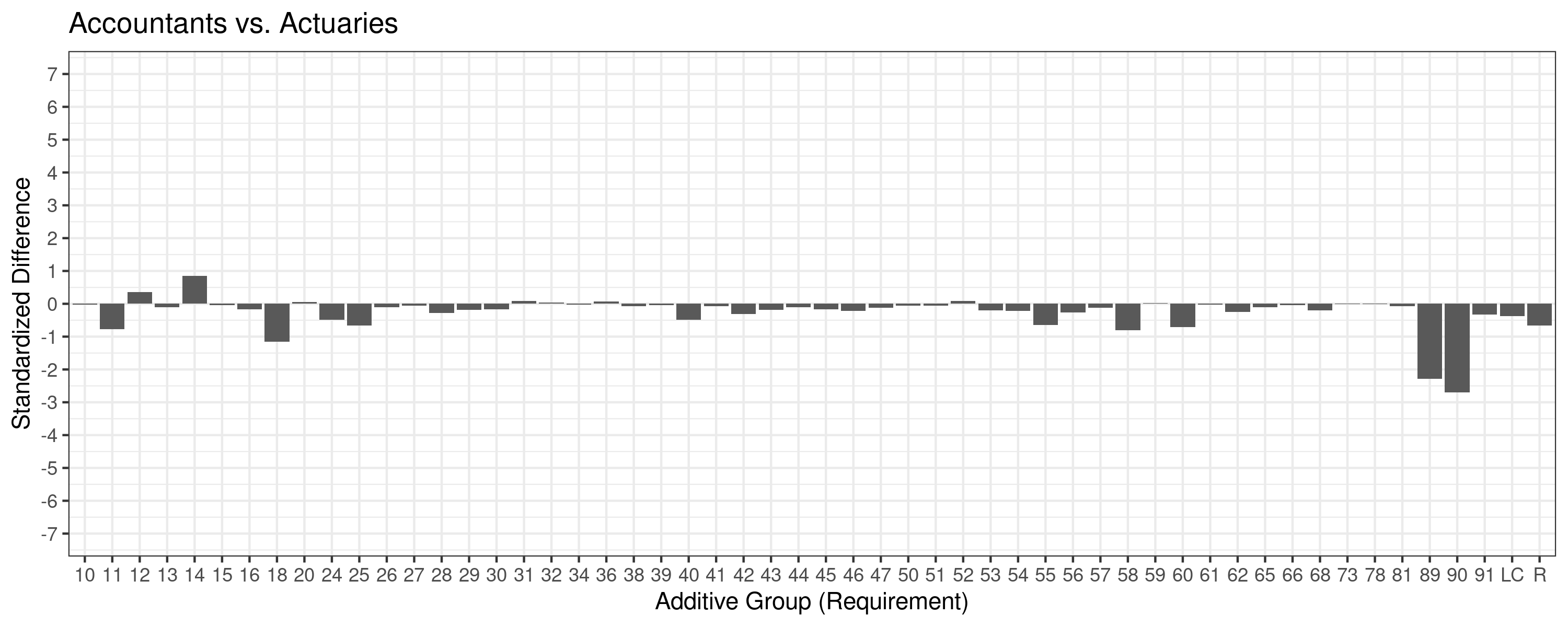}
    \end{measuredfigure}
    \caption{Difference in standardized ELEs between Accountants and Actuaries.}
    \label{fig:diff_acct-v-act}
\end{figure}

\begin{figure}[h]
    \centering
    \begin{measuredfigure}
    \includegraphics[height=6cm,width=15cm]{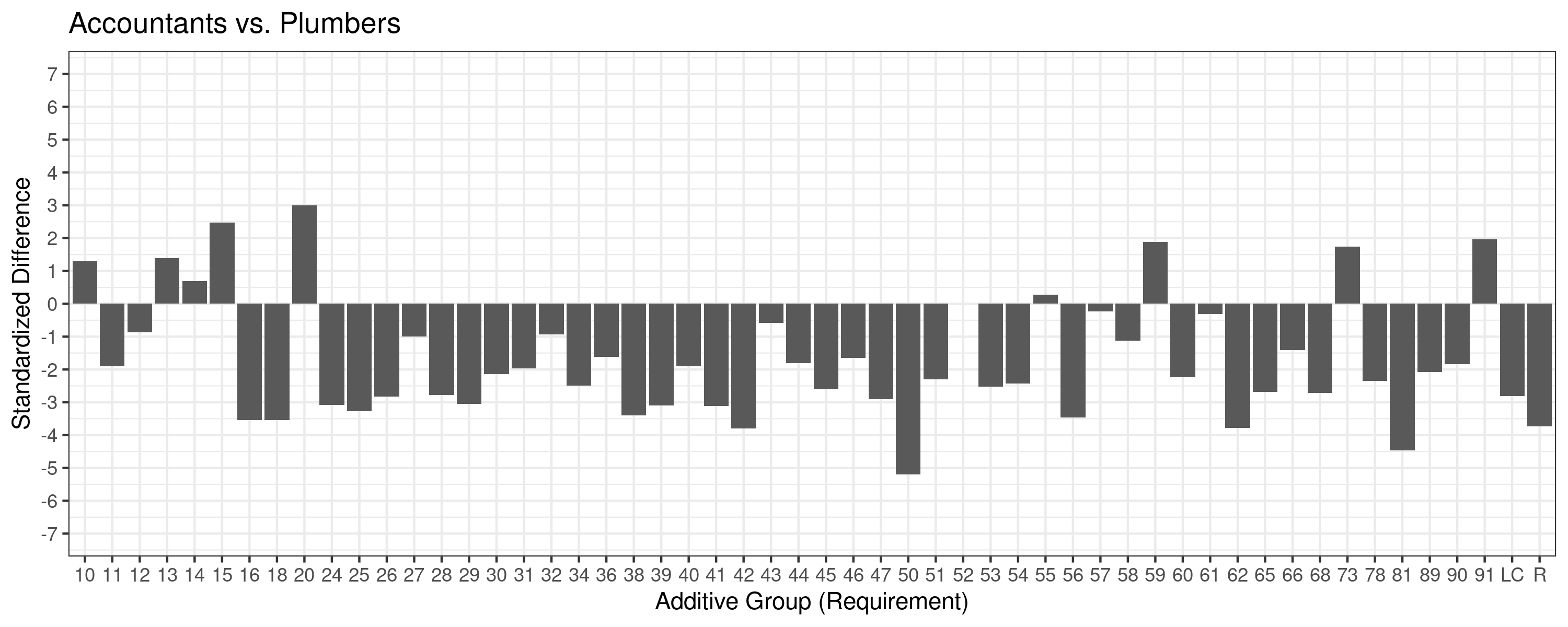}
    \end{measuredfigure}
    \caption{Difference in standardized ELEs between Accountants and Plumbers.}
    \label{fig:diff_acct-v-plumb}
\end{figure}

We also include figures demonstrating the differences between standardized expected values by occupation. Specifically, we visualize the same three pairs that were addressed in the overlap analysis, namely \textit{Legal Secretaries} vs. \textit{General Secretaries} (\figref{fig:diff_legSec-v-genSec}), \textit{Accountants} vs. \textit{Actuaries} (\figref{fig:diff_acct-v-act}), and \textit{Accountants} vs. \textit{Plumbers} (\figref{fig:diff_acct-v-plumb}). These standardized ELE values may also prove useful in determining job similarity, in addition to the correlation analysis of ELE described earlier (see \sectionref{subsubsec:mice-vs-model} and \sectionref{subsubsec:measuring-ELE}).

\subsection{WIGEM: A Generalized Imputation Method} \label{subsec:generalizability}

While our method was built specifically with the ORS in mind, we believe that it can be used in a more generalized fashion. Our method as is could be used to impute missing values in any sort of dataset following (or that can be transformed to follow) the same constraints as the ORS, namely that (a) all values are bounded on the interval [0,1], and further that (b) the data can be organized into distinct, hierarchical groups, and the sum of the values within each group must be 1. 

We abstract our method to an algorithm we call \textbf{WIGEM} (\textbf{W}eighted \textbf{I}terative \textbf{G}uess \textbf{E}stimation \textbf{M}ethod), that can be broken down into the following steps:

\begin{enumerate}
	\setlength\itemsep{0em}
	\item Throw out a handful of the known data using a k-folds approach. These observations constitute the mock missing data.
	\item Generate an initial prediction for the missing (actual and mock) data. This can be a naive guess, or a more nuanced method such as our smart guess procedure.
	\item Fit a model to the (weighted) data, using both the known and missing values. Known values should receive a weight of 1, while missing values (actual and mock) should receive weights of $< 1$. Use this fitted model to generate new predictions for the missing data (actual and mock).
	\item Fit a model on the newly generated predictions and updated (increased) weights for the missing values (actual and mock), while keeping the known values fixed at their actual values. Use this new model to update (generate new) predictions.
	\item Repeat step 4 until a metric of choice reaches a cutoff value. In our case, we calculated the RMSE of the mock missing values, and terminated the iteration phase when the difference in RMSE between subsequent iterations was $< 0.001$. One might also use $R^2$ or mean absolute error (MAE).
	\item Repeat steps 2-5, only this time without generating mock missing data (so that only the actual missing values are predicted). Iteration proceeds until the termination point (iteration number) determined in step 5. The predictions resulting from this model are the final predictions for the missing data.
	\item (Optional) Depending on availability of such data, errors associated with the (known) values being modeled can be used to generate simulations of the known data so that a distribution of estimates (and therefore confidence intervals) can be produced for each missing observation. If this step is possible, step 6 above will be applied to each of the simulations.
\end{enumerate}

Our method maximizes the use of the known data and its inherent constraints. Moreover, imputation is accomplished in an iterative fashion that terminates based on measurable errors. We believe WIGEM is a powerful algorithm for generating robust estimates for missing data.


Though we think this iterative guess methodology can be applied to other cases where a smart guess for the missing value is available but the values are unbounded, we believe survey modeling is the optimal application: the amount of missing information is bounded, the amount of known information is fixed and bounded by measured errors, and there are intuitive and measurable correlations between groups.

While the application of WIGEM described in this paper implemented a version of XGBoost that allows control of each input observation's contribution to the loss function using weights and limits, it should be noted any other model that gives the modeler control of the loss function contributions of prediction errors by observation would also work with WIGEM. Indeed, for the ambitious, many legacy fit algorithms have loss functions that can be accessed directly for this purpose.


\newpage


\bibliography{library}

\bibliographystyle{abbrv}

\newpage

\appendix
\section{Predictor Mapping} \label{app:A}

This Appendix contains a table describing the mapping used to produce the predictors \textbf{Requirement}, \textbf{Frequency}, \textbf{Intensity}, and \textbf{Requirement Category} from the original ORS data (as described in \sectionref{subsubsec:frequency} - \sectionref{subsubsec:req} and \sectionref{subsubsec:reqcat}). Note that both \textbf{Frequency} and \textbf{Intensity} were constructed such that all assigned values were on the interval [0,100]. The table starts on next page.

\begin{landscape}
    \tiny
    \begin{longtable}{p{5.5cm}llp{5.5cm}rrr}
    \textbf{\orsVarName{data\_element\_text}} & \textbf{\orsVarName{data\_type\_text}} & \textbf{\orsVarName{additive\_group}} & \textbf{Requirement} & \textbf{Frequency} & \textbf{Intensity} & \textbf{Category} \\
    \hline\endhead  
    \hline\endfoot  
         SVP & $>$ 1 MONTH, = 3 MONTHS & 10 & SVP & 100 & 4 & EDU \\
        SVP & $>$ 1 YEAR, = 2 YEARS & 10 & SVP & 100 & 20 & EDU \\
        SVP & $>$ 2 YEARS, = 4 YEARS & 10 & SVP & 100 & 40 & EDU \\
        SVP & $>$ 3 MONTHS, = 6 MONTHS & 10 & SVP & 100 & 5 & EDU \\
        SVP & $>$ 4 YEARS, = 10 YEARS & 10 & SVP & 100 & 60 & EDU \\
        SVP & $>$ 6 MONTHS, = 1 YEAR & 10 & SVP & 100 & 8 & EDU \\
        SVP & $>$SHORT DEMO, = 1 MONTH & 10 & SVP & 100 & 1 & EDU \\
        SVP & OVER 10 YEARS & 10 & SVP & 100 & 100 & EDU \\
        SVP & SHORT DEMO ONLY & 10 & SVP & 100 & 0 & EDU \\
        Literacy required & NO & 11 & Literacy required & 0 & 100 & EDU \\
        Literacy required & YES & 11 & Literacy required & 100 & 100 & EDU \\
        Pre-employment training & NO & 12 & Pre-employment training & 0 & 100 & EDU \\
        Pre-employment training & YES & 12 & Pre-employment training & 100 & 100 & EDU \\
        Prior work experience & NO & 13 & Prior work experience & 0 & 100 & EDU \\
        Prior work experience & YES & 13 & Prior work experience & 100 & 100 & EDU \\
        Post-employment training & NO & 14 & On the Job Training & 0 & 100 & EDU \\
        Post-employment training & YES & 14 & On the Job Training & 100 & 100 & EDU \\
        Sitting vs. standing/walking at will & NO & 15 & Sitting vs. standing/walking at will & 0 & 100 & PHY \\
        Sitting vs. standing/walking at will & YES & 15 & Sitting vs. standing/walking at will & 100 & 100 & PHY \\
        Reaching overhead & FREQUENTLY & 16 & Reaching & 67 & 100 & PHY \\
        Reaching overhead & NOT PRESENT & 16 & Reaching & 0 & 100 & PHY \\
        Reaching overhead & OCCASIONALLY & 16 & Reaching & 33 & 100 & PHY \\
        Reaching overhead & SELDOM & 16 & Reaching & 2 & 100 & PHY \\
        Reaching at/below the shoulder & CONSTANTLY & 18 & Reaching & 100 & 50 & PHY \\
        Reaching at/below the shoulder & FREQUENTLY & 18 & Reaching & 67 & 50 & PHY \\
        Reaching at/below the shoulder & NOT PRESENT & 18 & Reaching & 0 & 50 & PHY \\
        Reaching at/below the shoulder & OCCASIONALLY & 18 & Reaching & 33 & 50 & PHY \\
        Reaching at/below the shoulder & SELDOM & 18 & Reaching & 2 & 50 & PHY \\
        Keyboarding: Traditional & CONSTANTLY & 20 & Keyboarding: Traditional & 100 & 100 & PHY \\
        Keyboarding: Traditional & FREQUENTLY & 20 & Keyboarding: Traditional & 67 & 100 & PHY \\
        Keyboarding: Traditional & NOT PRESENT & 20 & Keyboarding: Traditional & 0 & 100 & PHY \\
        Keyboarding: Traditional & OCCASIONALLY & 20 & Keyboarding: Traditional & 33 & 100 & PHY \\
        Keyboarding: Traditional & SELDOM & 20 & Keyboarding: Traditional & 2 & 100 & PHY \\
        Lifting/carrying Seldom & $>$ 10 LBS, = 20 LBS & 24 & Lifting/carrying & 2 & 10 & PHY \\
        Lifting/carrying Seldom & $>$ 100 LBS & 24 & Lifting/carrying & 2 & 100 & PHY \\
        Lifting/carrying Seldom & $>$ 20 LBS, = 50 LBS & 24 & Lifting/carrying & 2 & 20 & PHY \\
        Lifting/carrying Seldom & $>$1 LBS, $<$=10 POUNDS & 24 & Lifting/carrying & 2 & 1 & PHY \\
        Lifting/carrying Seldom & $>$50 LBS, = 100 LBS & 24 & Lifting/carrying & 2 & 50 & PHY \\
        Lifting/carrying Seldom & NEGLIGIBLE & 24 & Lifting/carrying & 2 & 0.5 & PHY \\
        Lifting/carrying Seldom & NONE & 24 & Lifting/carrying & 2 & 0 & PHY \\
        Lifting/carrying Occasionally & $>$ 10 LBS, = 20 LBS & 25 & Lifting/carrying & 33 & 10 & PHY \\
        Lifting/carrying Occasionally & $>$ 20 LBS, = 50 LBS & 25 & Lifting/carrying & 33 & 20 & PHY \\
        Lifting/carrying Occasionally & $>$1 LBS, $<$=10 POUNDS & 25 & Lifting/carrying & 33 & 1 & PHY \\
        Lifting/carrying Occasionally & $>$50 LBS, = 100 LBS & 25 & Lifting/carrying & 33 & 50 & PHY \\
        Lifting/carrying Occasionally & NEGLIGIBLE & 25 & Lifting/carrying & 33 & 0.5 & PHY \\
        Lifting/carrying Occasionally & NONE & 25 & Lifting/carrying & 33 & 0 & PHY \\
        Lifting/carrying Frequently & $>$ 10 LBS, = 25 LBS & 26 & Lifting/carrying & 67 & 10 & PHY \\
        Lifting/carrying Frequently & $>$ 25 LBS, = 50 LBS & 26 & Lifting/carrying & 67 & 25 & PHY \\
        Lifting/carrying Frequently & $>$1 LBS, $<$=10 POUNDS & 26 & Lifting/carrying & 67 & 1 & PHY \\
        Lifting/carrying Frequently & NEGLIGIBLE & 26 & Lifting/carrying & 67 & 0.5 & PHY \\
        Lifting/carrying Frequently & NONE & 26 & Lifting/carrying & 67 & 0 & PHY \\
        Lifting/carrying Constantly & $>$ 10 LBS, = 20 LBS & 27 & Lifting/carrying & 100 & 10 & PHY \\
        Lifting/carrying Constantly & $>$ 20 LBS & 27 & Lifting/carrying & 100 & 20 & PHY \\
        Lifting/carrying Constantly & $>$1 LBS, $<$=10 POUNDS & 27 & Lifting/carrying & 100 & 1 & PHY \\
        Lifting/carrying Constantly & NEGLIGIBLE & 27 & Lifting/carrying & 100 & 0.5 & PHY \\
        Lifting/carrying Constantly & NONE & 27 & Lifting/carrying & 100 & 0 & PHY \\
        Crawling & NOT PRESENT & 28 & Crawling & 0 & 100 & PHY \\
        Crawling & OCCASIONALLY & 28 & Crawling & 33 & 100 & PHY \\
        Crawling & SELDOM & 28 & Crawling & 2 & 100 & PHY \\
        Crawling & CONSTANTLY & 28 & Crawling & 100 & 100 & PHY \\
        Crawling & FREQUENTLY & 28 & Crawling & 67 & 100 & PHY \\
        Pushing/pulling: Hands/arms & CONSTANTLY & 29 & Pushing/pulling: Hands/arms & 100 & 100 & PHY \\
        Pushing/pulling: Hands/arms & FREQUENTLY & 29 & Pushing/pulling: Hands/arms & 67 & 100 & PHY \\
        Pushing/pulling: Hands/arms & NOT PRESENT & 29 & Pushing/pulling: Hands/arms & 0 & 100 & PHY \\
        Pushing/pulling: Hands/arms & OCCASIONALLY & 29 & Pushing/pulling: Hands/arms & 33 & 100 & PHY \\
        Pushing/pulling: Hands/arms & SELDOM & 29 & Pushing/pulling: Hands/arms & 2 & 100 & PHY \\
        Pushing/pulling: Feet/legs & CONSTANTLY & 30 & Pushing/pulling: Feet/legs & 100 & 100 & PHY \\
        Pushing/pulling: Feet/legs & FREQUENTLY & 30 & Pushing/pulling: Feet/legs & 67 & 100 & PHY \\
        Pushing/pulling: Feet/legs & NOT PRESENT & 30 & Pushing/pulling: Feet/legs & 0 & 100 & PHY \\
        Pushing/pulling: Feet/legs & OCCASIONALLY & 30 & Pushing/pulling: Feet/legs & 33 & 100 & PHY \\
        Pushing/pulling: Feet/legs & SELDOM & 30 & Pushing/pulling: Feet/legs & 2 & 100 & PHY \\
        Driving & NO & 31 & Driving & 0 & 100 & PHY \\
        Driving & YES & 31 & Driving & 100 & 100 & PHY \\
        Foot/leg controls & CONSTANTLY & 32 & Foot/leg controls & 100 & 100 & PHY \\
        Foot/leg controls & FREQUENTLY & 32 & Foot/leg controls & 67 & 100 & PHY \\
        Foot/leg controls & NOT PRESENT & 32 & Foot/leg controls & 0 & 100 & PHY \\
        Foot/leg controls & OCCASIONALLY & 32 & Foot/leg controls & 33 & 100 & PHY \\
        Foot/leg controls & SELDOM & 32 & Foot/leg controls & 2 & 100 & PHY \\
        Gross manipulation & CONSTANTLY & 34 & Gross manipulation & 100 & 100 & PHY \\
        Gross manipulation & FREQUENTLY & 34 & Gross manipulation & 67 & 100 & PHY \\
        Gross manipulation & NOT PRESENT & 34 & Gross manipulation & 0 & 100 & PHY \\
        Gross manipulation & OCCASIONALLY & 34 & Gross manipulation & 33 & 100 & PHY \\
        Gross manipulation & SELDOM & 34 & Gross manipulation & 2 & 100 & PHY \\
        Fine manipulation & CONSTANTLY & 36 & Fine manipulation & 100 & 100 & PHY \\
        Fine manipulation & FREQUENTLY & 36 & Fine manipulation & 67 & 100 & PHY \\
        Fine manipulation & NOT PRESENT & 36 & Fine manipulation & 0 & 100 & PHY \\
        Fine manipulation & OCCASIONALLY & 36 & Fine manipulation & 33 & 100 & PHY \\
        Fine manipulation & SELDOM & 36 & Fine manipulation & 2 & 100 & PHY \\
        Climbing ramps or stairs (structure-related) & NO & 38 & Climbing ramps or stairs (structure-related) & 0 & 100 & PHY \\
        Climbing ramps or stairs (structure-related) & YES & 38 & Climbing ramps or stairs (structure-related) & 100 & 100 & PHY \\
        Climbing ramps or stairs (work-related) & NOT PRESENT & 39 & Climbing ramps or stairs (work-related) & 0 & 100 & PHY \\
        Climbing ramps or stairs (work-related) & OCCASIONALLY & 39 & Climbing ramps or stairs (work-related) & 33 & 100 & PHY \\
        Climbing ramps or stairs (work-related) & SELDOM & 39 & Climbing ramps or stairs (work-related) & 2 & 100 & PHY \\
        Climbing ramps or stairs (work-related) & CONSTANTLY & 39 & Climbing ramps or stairs (work-related) & 100 & 100 & PHY \\
        Climbing ramps or stairs (work-related) & FREQUENTLY & 39 & Climbing ramps or stairs (work-related) & 67 & 100 & PHY \\
        Climbing ladders, ropes, or scaffolds & NOT PRESENT & 40 & Climbing ladders, ropes, or scaffolds & 0 & 100 & PHY \\
        Climbing ladders, ropes, or scaffolds & OCCASIONALLY & 40 & Climbing ladders, ropes, or scaffolds & 33 & 100 & PHY \\
        Climbing ladders, ropes, or scaffolds & SELDOM & 40 & Climbing ladders, ropes, or scaffolds & 2 & 100 & PHY \\
        Climbing ladders, ropes, or scaffolds & CONSTANTLY & 40 & Climbing ladders, ropes, or scaffolds & 100 & 100 & PHY \\
        Climbing ladders, ropes, or scaffolds & FREQUENTLY & 40 & Climbing ladders, ropes, or scaffolds & 67 & 100 & PHY \\
        Stooping & FREQUENTLY & 41 & Stooping & 67 & 100 & PHY \\
        Stooping & NOT PRESENT & 41 & Stooping & 0 & 100 & PHY \\
        Stooping & OCCASIONALLY & 41 & Stooping & 33 & 100 & PHY \\
        Stooping & SELDOM & 41 & Stooping & 2 & 100 & PHY \\
        Stooping & CONSTANTLY & 41 & Stooping & 100 & 100 & PHY \\
        Kneeling & NOT PRESENT & 42 & Kneeling & 0 & 100 & PHY \\
        Kneeling & OCCASIONALLY & 42 & Kneeling & 33 & 100 & PHY \\
        Kneeling & SELDOM & 42 & Kneeling & 2 & 100 & PHY \\
        Kneeling & CONSTANTLY & 42 & Kneeling & 100 & 100 & PHY \\
        Kneeling & FREQUENTLY & 42 & Kneeling & 67 & 100 & PHY \\
        Extreme cold & NOT PRESENT & 43 & Extreme cold & 0 & 100 & ENV \\
        Extreme cold & OCCASIONALLY & 43 & Extreme cold & 33 & 100 & ENV \\
        Extreme cold & SELDOM & 43 & Extreme cold & 2 & 100 & ENV \\
        Extreme cold & CONSTANTLY & 43 & Extreme cold & 100 & 100 & ENV \\
        Extreme cold & FREQUENTLY & 43 & Extreme cold & 67 & 100 & ENV \\
        Extreme heat & CONSTANTLY & 44 & Extreme heat & 100 & 100 & ENV \\
        Extreme heat & FREQUENTLY & 44 & Extreme heat & 67 & 100 & ENV \\
        Extreme heat & NOT PRESENT & 44 & Extreme heat & 0 & 100 & ENV \\
        Extreme heat & OCCASIONALLY & 44 & Extreme heat & 33 & 100 & ENV \\
        Extreme heat & SELDOM & 44 & Extreme heat & 2 & 100 & ENV \\
        Wetness & CONSTANTLY & 45 & Wetness & 100 & 100 & ENV \\
        Wetness & FREQUENTLY & 45 & Wetness & 67 & 100 & ENV \\
        Wetness & NOT PRESENT & 45 & Wetness & 0 & 100 & ENV \\
        Wetness & OCCASIONALLY & 45 & Wetness & 33 & 100 & ENV \\
        Wetness & SELDOM & 45 & Wetness & 2 & 100 & ENV \\
        Humidity & FREQUENTLY & 46 & Humidity & 67 & 100 & ENV \\
        Humidity & NOT PRESENT & 46 & Humidity & 0 & 100 & ENV \\
        Humidity & OCCASIONALLY & 46 & Humidity & 33 & 100 & ENV \\
        Humidity & SELDOM & 46 & Humidity & 2 & 100 & ENV \\
        Humidity & CONSTANTLY & 46 & Humidity & 100 & 100 & ENV \\
        Heavy vibrations & NOT PRESENT & 47 & Heavy vibrations & 0 & 100 & ENV \\
        Heavy vibrations & OCCASIONALLY & 47 & Heavy vibrations & 33 & 100 & ENV \\
        Heavy vibrations & CONSTANTLY & 47 & Heavy vibrations & 100 & 100 & ENV \\
        Heavy vibrations & FREQUENTLY & 47 & Heavy vibrations & 67 & 100 & ENV \\
        Heavy vibrations & SELDOM & 47 & Heavy vibrations & 2 & 100 & ENV \\
        High, exposed places & NOT PRESENT & 50 & High, exposed places & 0 & 100 & ENV \\
        High, exposed places & OCCASIONALLY & 50 & High, exposed places & 33 & 100 & ENV \\
        High, exposed places & SELDOM & 50 & High, exposed places & 2 & 100 & ENV \\
        Proximity to moving mechanical parts & CONSTANTLY & 51 & Proximity to moving mechanical parts & 100 & 100 & ENV \\
        Proximity to moving mechanical parts & FREQUENTLY & 51 & Proximity to moving mechanical parts & 67 & 100 & ENV \\
        Proximity to moving mechanical parts & NOT PRESENT & 51 & Proximity to moving mechanical parts & 0 & 100 & ENV \\
        Proximity to moving mechanical parts & OCCASIONALLY & 51 & Proximity to moving mechanical parts & 33 & 100 & ENV \\
        Proximity to moving mechanical parts & SELDOM & 51 & Proximity to moving mechanical parts & 2 & 100 & ENV \\
        Near visual acuity & NO & 52 & Near visual acuity & 0 & 100 & PHY \\
        Near visual acuity & YES & 52 & Near visual acuity & 100 & 100 & PHY \\
        Far visual acuity & NO & 53 & Far visual acuity & 0 & 100 & PHY \\
        Far visual acuity & YES & 53 & Far visual acuity & 100 & 100 & PHY \\
        Peripheral vision & NO & 54 & Peripheral vision & 0 & 100 & PHY \\
        Peripheral vision & YES & 54 & Peripheral vision & 100 & 100 & PHY \\
        Communicating verbally & CONSTANTLY & 55 & Speaking & 100 & 100 & PHY \\
        Communicating verbally & FREQUENTLY & 55 & Speaking & 67 & 100 & PHY \\
        Communicating verbally & NOT PRESENT & 55 & Speaking & 0 & 100 & PHY \\
        Communicating verbally & OCCASIONALLY & 55 & Speaking & 33 & 100 & PHY \\
        Communicating verbally & SELDOM & 55 & Speaking & 2 & 100 & PHY \\
        Noise intensity level & LOUD & 56 & Noise intensity level & 100 & 75 & ENV \\
        Noise intensity level & MODERATE & 56 & Noise intensity level & 100 & 40 & ENV \\
        Noise intensity level & QUIET & 56 & Noise intensity level & 100 & 0 & ENV \\
        Hearing requirements: One-on-one & NO & 57 & Hearing requirements: One-on-one & 0 & 100 & PHY \\
        Hearing requirements: One-on-one & YES & 57 & Hearing requirements: One-on-one & 100 & 100 & PHY \\
        Hearing requirements: Group or conference & NO & 58 & Hearing requirements: Group or conference & 0 & 100 & PHY \\
        Hearing requirements: Group or conference & YES & 58 & Hearing requirements: Group or conference & 100 & 100 & PHY \\
        Hearing requirements: Telephone & NO & 59 & Hearing requirements: Telephone & 0 & 100 & PHY \\
        Hearing requirements: Telephone & YES & 59 & Hearing requirements: Telephone & 100 & 100 & PHY \\
        Hearing requirements: Other Sounds & NO & 60 & Hearing requirements: Other Sounds & 0 & 100 & PHY \\
        Hearing requirements: Other Sounds & YES & 60 & Hearing requirements: Other Sounds & 100 & 100 & PHY \\
        Hearing requirements: Pass a hearing test & NO & 61 & Hearing requirements: Pass a hearing test & 0 & 100 & PHY \\
        Hearing requirements: Pass a hearing test & YES & 61 & Hearing requirements: Pass a hearing test & 100 & 100 & PHY \\
        Crouching & FREQUENTLY & 62 & Crouching & 67 & 100 & PHY \\
        Crouching & NOT PRESENT & 62 & Crouching & 0 & 100 & PHY \\
        Crouching & OCCASIONALLY & 62 & Crouching & 33 & 100 & PHY \\
        Crouching & SELDOM & 62 & Crouching & 2 & 100 & PHY \\
        Outdoors & CONSTANTLY & 65 & Outdoors & 100 & 100 & ENV \\
        Outdoors & FREQUENTLY & 65 & Outdoors & 67 & 100 & ENV \\
        Outdoors & NOT PRESENT & 65 & Outdoors & 0 & 100 & ENV \\
        Outdoors & OCCASIONALLY & 65 & Outdoors & 33 & 100 & ENV \\
        Outdoors & SELDOM & 65 & Outdoors & 2 & 100 & ENV \\
        Pushing/pulling: Feet only & CONSTANTLY & 66 & Pushing/pulling: feet only & 100 & 100 & PHY \\
        Pushing/pulling: Feet only & NOT PRESENT & 66 & Pushing/pulling: feet only & 0 & 100 & PHY \\
        Pushing/pulling: Feet only & OCCASIONALLY & 66 & Pushing/pulling: feet only & 33 & 100 & PHY \\
        Strength & HEAVY WORK & 68 & Strength & 100 & 70 & PHY \\
        Strength & LIGHT WORK & 68 & Strength & 100 & 5 & PHY \\
        Strength & MEDIUM WORK & 68 & Strength & 100 & 25 & PHY \\
        Strength & SEDENTARY & 68 & Strength & 100 & 0 & PHY \\
        Strength & VERY HEAVY WORK & 68 & Strength & 100 & 100 & PHY \\
        Minimum education level & ASSOCIATE'S & 73 & Minimum education level & 100 & 35 & EDU \\
        Minimum education level & ASSOCIATE'S VOCATIONAL & 73 & Minimum education level & 100 & 40 & EDU \\
        Minimum education level & BACHELOR'S & 73 & Minimum education level & 100 & 50 & EDU \\
        Minimum education level & DOCTORATE & 73 & Minimum education level & 100 & 100 & EDU \\
        Minimum education level & HIGH SCHOOL & 73 & Minimum education level & 100 & 20 & EDU \\
        Minimum education level & HIGH SCHOOL VOCATIONAL & 73 & Minimum education level & 100 & 25 & EDU \\
        Minimum education level & MASTER'S & 73 & Minimum education level & 100 & 70 & EDU \\
        Minimum education level & NONE & 73 & Minimum education level & 100 & 0 & EDU \\
        Minimum education level & PROFESSIONAL & 73 & Minimum education level & 100 & 80 & EDU \\
        Sitting & SIT & 78 & Sitting or standing/walking & 100 & 50 & PHY \\
        Standing/walking & STAND & 78 & Sitting or standing/walking & 100 & 100 & PHY \\
        Hazardous contaminants & NOT PRESENT & 81 & Hazardous contaminants & 0 & 100 & ENV \\
        Hazardous contaminants & OCCASIONALLY & 81 & Hazardous contaminants & 33 & 100 & ENV \\
        Hazardous contaminants & SELDOM & 81 & Hazardous contaminants & 2 & 100 & ENV \\
        Pre-employment training: Certification & NO & 89 & Pre-employment training: Certification & 0 & 100 & EDU \\
        Pre-employment training: Certification & YES & 89 & Pre-employment training: Certification & 100 & 100 & EDU \\
        Pre-employment training: License & NO & 90 & Pre-employment training: License & 0 & 100 & EDU \\
        Pre-employment training: License & YES & 90 & Pre-employment training: License & 100 & 100 & EDU \\
        Pre-employment training: Educational Certificate & NO & 91 & Pre-employment training: Educational Certificate & 0 & 100 & EDU \\
        Pre-employment training: Educational Certificate & YES & 91 & Pre-employment training: Educational Certificate & 100 & 100 & EDU \\
    \label{tab:appA}
    \end{longtable}
\end{landscape}

\end{document}